\def\isextended{1}
\def\isdraft{1}

\ifdefined\isdraft
\documentclass[12pt,onecolumn,draftclsnofoot]{IEEEtran}
\else
\documentclass[10pt]{IEEEtran}
\fi

\usepackage{amsmath,amsfonts,amssymb}
\usepackage{amsthm}
\usepackage{graphicx}
\usepackage{algorithm,algorithmicx,algpseudocode}
\usepackage{color}

\newtheorem{theorem}{Theorem}
\newtheorem{lemma}{Lemma}
\newtheorem{proposition}{Proposition}

\newtheorem{example}{Example}

\theoremstyle{definition}
\newtheorem{definition}{Definition}

\ifdefined\isreview
	\newcommand{\addchange}[1]{{\color{blue}#1}}
\else
	\newcommand{\addchange}[1]{#1}
\fi

\newcommand{\appref}[1]{\addchange{Appendix~\ref{#1}}}

\newcommand{\apprefthm}[3]{#2~\ref{#3} in Appendix~\ref{#1}}
\ifdefined\isextended
	\newcommand{\apprefextra}{Appendix~\ref{app:extra}}
\else
	\newcommand{\apprefextra}{\addchange{\cite[Appendix~F]{extended}}}
\fi

\newcommand{\floor}[1]{\left\lfloor#1\right\rfloor}
\newcommand{\ceil}[1]{\left\lceil#1\right\rceil}

\graphicspath{ {./figures/} }

\IEEEoverridecommandlockouts

\newcommand{\myplotwidth}{.4\textwidth}
\newcommand{\myfigscale}{.45}

\begin{document}

\title{Caching with Partial Adaptive Matching}

\ifdefined\isextended
\author{Jad Hachem, \and
Nikhil Karamchandani, \and
Sharayu Moharir, \and
Suhas Diggavi%
\thanks{Shorter versions of the results in this paper have appeared in IEEE ISIT 2017~\cite{HKMDuniform} and IEEE ITW 2017~\cite{HKMDzipf}.

J. Hachem and S. Diggavi's research was supported in part by NSF grants \#1423271 and \#1514531.

N. Karamchandani's research was supported in part by Indo-French grant No. IFC/DST-Inria-2016-01/448 ``Machine Learning for Network Analytics'' and a seed grant from IIT Bombay.

S. Moharir's research was supported in part by a seed grant from IIT Bombay.}}
\else
\author{Jad Hachem, \IEEEmembership{Student Member}, \and
Nikhil Karamchandani, \IEEEmembership{Member}, \and
Sharayu Moharir, \IEEEmembership{Member}, \and
Suhas Diggavi, \IEEEmembership{Fellow}%
\thanks{Shorter versions of the results in this paper have appeared in IEEE ISIT 2017~\cite{HKMDuniform} and IEEE ITW 2017~\cite{HKMDzipf}.

J. Hachem and S. Diggavi's research was supported in part by NSF grants \#1423271 and \#1514531.

N. Karamchandani's research was supported in part by Indo-French grant No. IFC/DST-Inria-2016-01/448 ``Machine Learning for Network Analytics'' and a seed grant from IIT Bombay.

S. Moharir's research was supported in part by a seed grant from IIT Bombay.}}
\fi

\maketitle

\begin{abstract}
We study the caching problem when we are allowed to match each user to one of a subset of caches after its request is revealed.
We focus on non-uniformly popular content, specifically when the file popularities obey a Zipf distribution.
We study two extremal schemes, one focusing on coded server transmissions while ignoring matching capabilities, and the other focusing on adaptive matching while ignoring potential coding opportunities.
We derive the rates achieved by these schemes and characterize the regimes in which one outperforms the other.
We also compare them to information-theoretic outer bounds, and finally propose a hybrid scheme that generalizes ideas from the two schemes and performs at least as well as either of them in most memory regimes.

\end{abstract}

\ifdefined\isextended
\else
\fi

\section{Introduction}
\label{sec:intro}

In modern content distribution networks, caching is a technique that places popular content at nodes close to the end users in order to reduce the overall network traffic.
In \cite{maddah-ali2012}, a new ``coded caching'' technique was introduced for broadcast networks.
This technique places different content in each cache, and takes advantage of these differences to send a common coded broadcast message to multiple users at once.
This was shown not only to greatly reduce the network load in comparison with traditional uncoded techniques, but also to be approximately optimal in general.

\addchange{The problem is motivated by wireless heterogeneous networks, which consist of a dense deployment of wireless access points (e.g., small cells) with short range but high communication rates, combined with a sparse deployment of base stations with long range but limited rate.
By equipping the access points with caches, users can potentially connect to one of several access points and gain access to their caches, while a base station can transmit a broadcast signal to many users at once in order to help serve their content requests \cite{HKDmultilevel}.}

In \cite{maddah-ali2012} as well as many other works in the literature \cite{niesen2013,Zcodedcaching,ZhangArbitrary,HKDmultilevel,MLcodedcaching}, a key assumption is that users are pre-fixed to specific caches; see also \cite{MNCommMag,PBLCDCommMag} for a survey of related works.
More precisely, each user connects to a specific cache before it requests a file from the content library.
This assumption was relaxed in \cite{leconte2012,moharir2016} where the system is allowed to choose a matching of users to caches \emph{after} the users make their requests, while respecting a per-cache load constraint.
In particular, after each user requests a file, any user could be matched to any cache as long as no cache had more than one user connected to it.
In this \emph{adaptive matching} setup, it was shown under certain request distributions that a coded delivery, while approximately optimal in the pre-fixed matching case, is unnecessary.
Indeed, it is sufficient to simply store complete files in the caches, and either connect a user to a cache containing its file or directly serve it from the server.

The above dichotomy indicates a fundamental difference between the system with completely pre-fixed matching and the system with full adaptive matching.
In this paper, we take a first step towards bridging the gap between these two extremes.
Our contributions are the following.

\begin{figure}
\centering
\includegraphics[width=.48\textwidth]{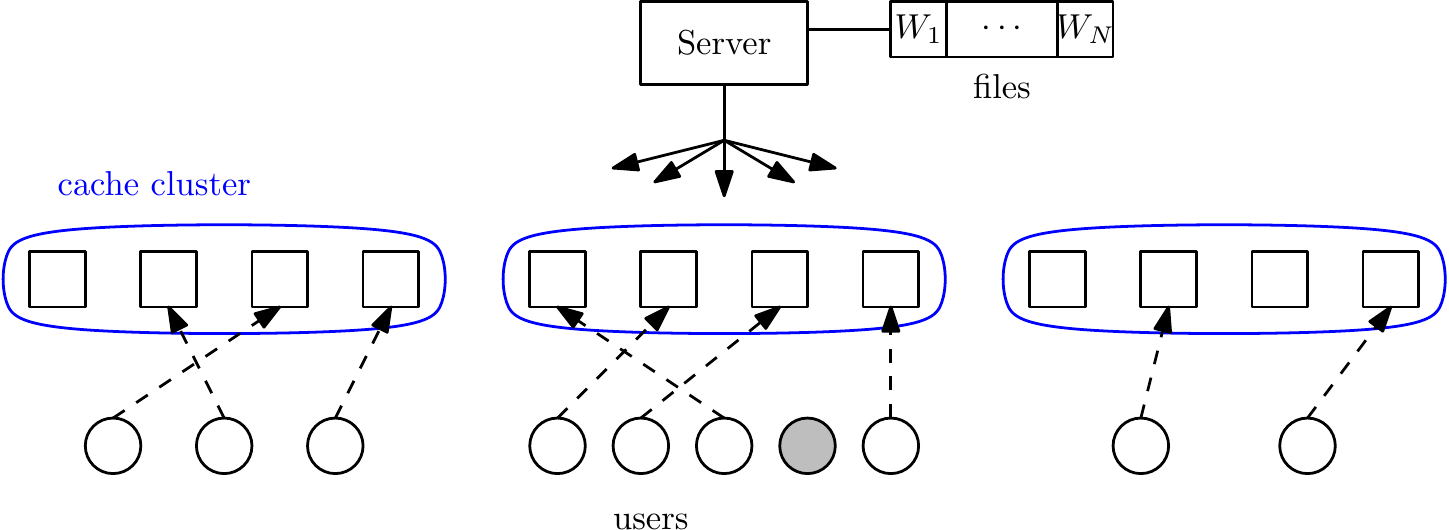}
\caption{Illustration of the setup considered in this paper.
The squares represent $K=12$ caches, divided into three clusters of size $d=4$ caches each, and the circles represent users at these clusters.
Dashed arrows represent the matching phase, and solid arrows the delivery phase.
Unmatched users are in gray.}
\label{fig:setup}
\end{figure}

We introduce a ``partial adaptive matching'' setup in which users can be matched to any cache belonging to a \emph{subset} of caches.
This can arise when \addchange{users do not have global reach.}
For instance, only some \addchange{access points in a wireless heterogeneous network} would be close enough to a user to ensure a potential reliable connection.
To make matters simple, \addchange{we model this by assuming} that the caches are partitioned into \emph{clusters} of fixed and equal size, and each user can be matched to any cache within a single cluster, as illustrated in \figurename~\ref{fig:setup} and described precisely in Section~\ref{sec:setup}.
This setup generalizes both setups considered above: on one extreme, if each cluster consisted of only a single cache, then the setup becomes the pre-fixed matching setup of \cite{maddah-ali2012}; on the other extreme, if all caches belonged to a single cluster, then we get back the total adaptive matching setup from \cite{leconte2012,moharir2016}.

Through this setup, we observe a dichotomy between coded caching and adaptive matching by considering two schemes: one scheme exclusively uses coded caching without taking advantage of adaptive matching; the other scheme focuses only on adaptive matching and performs uncoded delivery.
We show that the scheme prioritizing coded caching is more efficient when both the cache memory and the cluster size are small, while the scheme prioritizing adaptive matching is more efficient in the opposite case (see Theorems~\ref{thm:regimes-01} and~\ref{thm:regimes-12}).

We derive information-theoretic cut-set bounds on the optimal expected rate.
These bounds show that the coded caching scheme is approximately optimal in the small memory regime and adaptive matching is approximately optimal in the large memory regime (see Theorem~\ref{thm:converse}).

For a subclass of file popularities, we propose a new hybrid scheme that combines ideas from both the coded delivery and the adaptive matching schemes, and that performs better than both in most memory regimes (see Theorem~\ref{thm:hcm}).

In this paper, we consider the relevant case when some files are much more popular than others.
In particular, we consider a class of popularity distributions where file popularity follows a power law, specifically a Zipf law \cite{breslau1999web}.
This includes the special case when all files are equally popular, but also encompasses more skewed popularities.
We observe a difference in the behavior of the schemes for two subclasses of Zipf distributions, namely when the Zipf popularities are heavily skewed (the \emph{steep Zipf} case) and when they are not heavily skewed (the \emph{shallow Zipf} case).

The rest of this paper is organized as follows.
Section~\ref{sec:setup} precisely describes the problem setup and introduces the clustering model.
Section~\ref{sec:prelim} provides a preliminary discussion that prepares for the main results.
We divide the main results into Section~\ref{sec:shallow}, which focuses on the shallow Zipf case, and Section~\ref{sec:steep}, which focuses on the steep Zipf case.
Detailed proofs are given in the appendices.
\ifdefined\isextended\else
For lack of space, a few of the proofs are given in the extended version of this submission \cite{extended}, in Appendix~F.
\fi

\section{Problem Setup}
\label{sec:setup}

Consider the \addchange{centralized caching} system depicted in \figurename~\ref{fig:setup}.
A server holds $N$ files $W_1,\ldots,W_N$ of size $F$ bits each.
There are $K$ caches of capacity $MF$ bits, equivalently $M$ files, each.
The caches are divided into $K/d$ clusters of size $d$ each, where $d$ is assumed to divide $K$.
For every $n\in\{1,\ldots,N\}$ and every $c\in\{1,\ldots,K/d\}$, there are $u_n(c)$ users accessing cluster $c$ and requesting file $W_n$.
We refer to the numbers $\{u_n(c)\}_{n,c}$ as the \emph{request profile} and will often represent the request profile as a vector $\mathbf{u}$ for convenience.

In this paper, we focus on the case where the numbers $u_n(c)$ are independent Poisson random variables with parameter $\rho dp_n$, where $\rho\in(0,1/2)$ is some fixed constant%
\addchange{\footnote{\addchange{The restriction $\rho<1/2$ is for technical reasons and simplifies much of the analysis.
We believe our results should generalize to any $\rho<1$.}}}
and $p_1,\ldots,p_N$ is the \emph{popularity distribution} of the files, with $p_n\ge0$ and $p_1+\cdots+p_N=1$.
Thus $p_n$ represents the probability that a fixed user will request file $W_n$.
We particularly focus on the case where the files follow a Zipf law, i.e., $p_n\propto n^{-\beta}$ where $\beta\ge0$ is the Zipf parameter.
Note that the expected total number of users in the system is $\rho K$.

\addchange{The system operates in \emph{three phases}: in addition to the usual placement and delivery phases common to standard coded caching setups \cite{maddah-ali2012}, there is an intermediate phase that we call the \emph{matching phase}.
The placement and delivery phases have already been covered in the literature \cite{maddah-ali2012} and we here emphasize the matching phase.
In the first phase (placement), content related to the files is placed in the caches.
In the second phase (matching),} each user is matched to a single cache \emph{within its cluster}, with the constraint that no more than one user can be matched to a cache.%
\addchange{\footnote{\addchange{This load constraint on the caches is motivated by current systems which are limited to point-to-point communication and an underlying scheduling scheme.
While current systems also use point-to-point communication with the base stations, this paper relaxes this assumption for the base stations first, which are the communication bottleneck.
}}}
If there are fewer caches than users in one cluster, then some users will be unmatched.
\addchange{In the third phase (delivery), each user requests a file, and a common broadcast message is sent to all users.
Each user uses the message, along with the contents of its cache if it was matched to any, to recover its requested file.}

For a given request profile $\mathbf{u}$, let $R_{\mathbf{u}}$ denote the rate of the broadcast message required to deliver to all users their requested files.
For any cache memory $M$, our goal is to minimize the expected rate $\bar R=\mathbb{E}_{\mathbf{u}}[R_{\mathbf{u}}]$.
Specifically, we are interested in $\bar R^*$ defined as the smallest $\bar R$ over all possible strategies.
Furthermore, we assume that there are more files than caches, i.e., $N\ge K$, which is the case of most interest.
We also, for analytical convenience, focus on the case where the cluster size $d$ grows at least as fast as $\log K$.
More precisely, we assume
\begin{equation}
\label{eq:d-vs-K}
d \ge \left[ 2(1+t_0)/\alpha \right] \log K,
\end{equation}
where $\alpha=-\log(2\rho e^{1-2\rho})$ and $t_0>0$ is some constant.
Note that $\alpha>0$.
Other than analytical convenience, the reason for such a lower bound on $d$ is that, when $d$ is too small, the Poisson request model adopted in this paper is no longer suitable.
Indeed, if for example $d=1$, then with high probability a significant fraction of users will not be matched to any cache, leading to a rate proportional to $K$ even with infinite cache memory.

Finally, we will frequently use the helpful notation $[x]^+=\max\{x,0\}$ for all real numbers $x$.

\section{Preliminary Discussion}
\label{sec:prelim}

The setup we consider is a generalization of the pre-fixed matching setup (when $d=1$) and the maximal adaptive matching setup (when $d=K$).
From the literature, we know that different strategies are required for these two extremes: one using a coded delivery when $d=1$, and one using adaptive matching when $d=K$.%
\footnote{The request model used in the literature when $d=1$ is usually not the Poisson model used here.
 Instead, a multinomial model is used in which the total number of users is always fixed.
 As mentioned at the end of Section~\ref{sec:setup}, the Poisson model is not suitable in that case.
 However, the results from the literature are still very relevant to this paper.}
Therefore, there must be some transition in the suitable strategy as the cluster size $d$ increases from one to $K$.

\begin{figure}
\centering
\includegraphics[width=.35\textwidth]{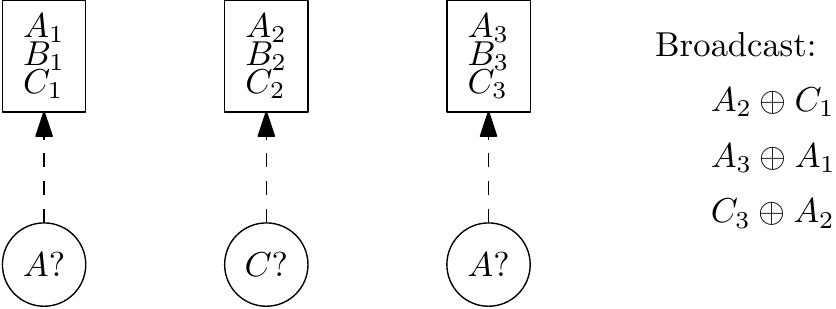}
\caption{\addchange{Illustration of PCD.
With PCD, the emphasis is on creating coding opportunities for any demand profile at the cost of ignoring adaptive matching.
We split each file $W$ into three equal parts ($W_1,W_2,W_3$), each of which is stored in one cache.
Thus the placement is perfectly symmetric with respect to the files and the caches.
When users join and request files, they are arbitrarily matched to caches (all matchings are equivalent because of the symmetry) and coded messages are broadcast to satisfy their demands.}}
\label{fig:pcd-example}
\end{figure}

\begin{figure}
\centering
\includegraphics[width=.35\textwidth]{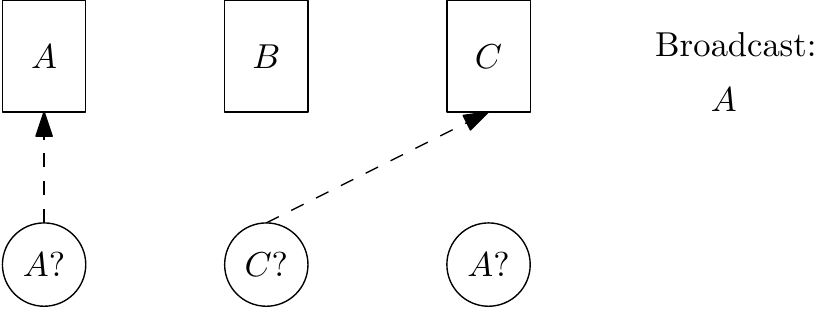}
\caption{\addchange{Illustration of PAM.
With PAM, the emphasis is on matching users to caches that contain their wanted content: the caches store complete files while the server directly serve users who could not be matched to a cache, ignoring the possibility of coding opportunities during delivery.
In this example, there is no matching that will locally serve all the users, and so at least one user must be served by the broadcast (in this case, the user on the right).}}
\label{fig:pam-example}
\end{figure}

The goal of this paper is to study this transition.
To do that, we first adapt and apply the strategies suitable for the two extremes to our intermediate case.
These strategies will exclusively focus on one of coded delivery and adaptive matching, and we will hence refer to them as ``Pure Coded Delivery'' (PCD) and ``Pure Adaptive Matching'' (PAM).
In particular, PCD will perform an arbitrary matching and apply the coded caching scheme from \cite{Zcodedcaching,ZhangArbitrary}, whereas PAM will apply a matching scheme similar to \cite{leconte2012,moharir2016} independently on each cluster and serve unmatched requests directly, ignoring any coding opportunities.
We then compare PCD and PAM in various regimes and evaluate them against information-theoretic outer bounds.
We illustrate the core idea of each scheme with the following example.
\addchange{%
\begin{example}
Suppose there are $K=3$ caches and $N=3$ equally popular files ($p_1=p_2=p_3=1/3$), called $A$, $B$, and $C$.
Suppose also that each cache can store up to one file ($M=1$), and that the caches form one cluster so that any user can be matched to any cache.
\figurename~\ref{fig:pcd-example} illustrates PCD in this situation and \figurename~\ref{fig:pam-example} illustrates PAM.
\end{example}
}

Regardless of the value $\beta$ of the Zipf parameter, we find that PCD tends to perform better than PAM when the cache memory $M$ is small, while PAM is superior to PCD when $M$ is large.
The particular threshold of $M$ where PAM overtakes PCD obeys an inverse relation with the cluster size $d$.
Thus when $d$ is small, PCD is the better choice for most memory values, whereas when $d$ is large, PAM performs better for most memory values.
This observation agrees with previous results on the two extremes $d=1$ and $d=K$, and it is illustrated in \figurename~\ref{fig:rate-01} and~\ref{fig:rate-12} and made precise in the theorems that follow.

\addchange{In this paper, most of the results pose no restrictions on the parameters except for \eqref{eq:d-vs-K} and $N\ge K$.
However, for some discussions and results (Theorem~\ref{thm:pam-12} in particular), we focus on the regime where $K$ grows (and thus so do $d$ and $N$), and asymptotic notation is to be understood with respect to the growth of $K$.}

\addchange{In addition, it will sometimes be useful to}
compare PCD and PAM under the restriction that the parameters all scale as powers of $K$.
This \addchange{is not assumed in the results but} can provide some high-level insights into \addchange{and visualization of} the different regimes where PCD or PAM dominate, while ignoring sub-polynomial factors such as $\log N$, thus simplifying the analysis.
During this polynomial-scaling-with-$K$ analysis---which we will call \emph{poly-$K$ analysis} for short---we will assume that
\begin{IEEEeqnarray*}{rCl'rCl'rCl}
N &=& K^\nu; & d &=& K^\delta; & M &=& K^\mu,\\
\nu &\ge& 1, & \delta &\in& (0,1], & \mu &\in& [0,1].
\IEEEyesnumber\label{eq:polyk}
\label{eq:polyk-n}
\label{eq:polyk-d}
\label{eq:polyk-m}
\end{IEEEeqnarray*}
We stress again that our results do not make these assumptions, but that this representation provides a useful way to \addchange{visualize} the results.

To proceed, we will separately consider two regimes for the Zipf popularity: a \emph{shallow Zipf} case in which $\beta\in[0,1)$, and a \emph{steep Zipf} case where $\beta>1$.%
\footnote{The case $\beta=1$ is a special case that usually requires separate handling.
We skip it in this paper, and analyzing it is part of our on-going work.}

\section{The Shallow Zipf Case ($\beta<1$)}
\label{sec:shallow}

\subsection{Comparing PCD and PAM when $\beta<1$}

The next theorem gives the rate achieved by PCD.

\begin{theorem}
\label{thm:pcd-01}
When $\beta\in[0,1)$, the PCD scheme can achieve for all $M$ an expected rate of
\[
\bar R^\mathrm{PCD} = \min\left\{ \rho K,
\left[ \tfrac{N}{M} - 1 \right]^+ + \tfrac{K^{-t_0}}{\sqrt{2\pi}} \right\}.
\]
\end{theorem}

Theorem~\ref{thm:pcd-01} can be proved by directly applying any suitable coded caching strategy \cite{Zcodedcaching,ZhangArbitrary,HKDmultilevel} along with an arbitrary matching phase.
The additional $K^{-t_0}$ term represents the expected number of users that will not be matched to any cache and must hence be served directly from the server.
The derivation of this term is done in \apprefthm{app:unmatched}{Lemma}{lemma:pcd-unmatched}.
\addchange{An interesting aspect of PCD is that, in order to maximize its coding opportunities, it treats all files as though they have equal popularity, at least in the shallow Zipf case.}

The next theorem gives the rate achieved by PAM.

\begin{theorem}
\label{thm:pam-01}
When $\beta\in[0,1)$, the PAM scheme can achieve an expected rate of
\[
\bar R^\mathrm{PAM} = \begin{cases}
\rho K & \text{if \addchange{$M<N/(1-\beta)d$};}\\
\min\left\{ \rho K, KMe^{-zdM/N} \right\} & \text{if \addchange{$M\ge N/(1-\beta)d$},}
\end{cases}
\]
where $z=(1-\beta)\rho h( (1+\rho)/2\rho ) >0$ with $h(x)=x\log x+1-x$.
\end{theorem}

Theorem~\ref{thm:pam-01} can be proved using a similar argument to \cite{leconte2012}: the idea is to replicate each file across the caches in each cluster, and match each user to a cache containing its requested file.
\addchange{Users that cannot be matched to a cache containing their file must be served directly by the server.
Contrary to PCD, which in the shallow Zipf case treats all files as equally likely, PAM leverages popularity by storing the more popular files in a larger number of caches.}
The detailed proof is given in \appref{app:pam-01}.
Notice that PAM can achieve a rate of $o(1)$ when $dM>\Omega(N\log N)$.%
\addchange{\footnote{\addchange{%
Recall that we have imposed a service constraint of one user per cache in our setup.
If we instead allow multiple users to access the same cache, then it can be shown that a rate of $o(1)$ can be achieved if and only if $dM>(1-o(1))N$.
Consequently, the cache service constraint increases this memory threshold by at most a logarithmic factor.
}}}

\begin{figure}
\centering
\includegraphics[width=\myplotwidth]{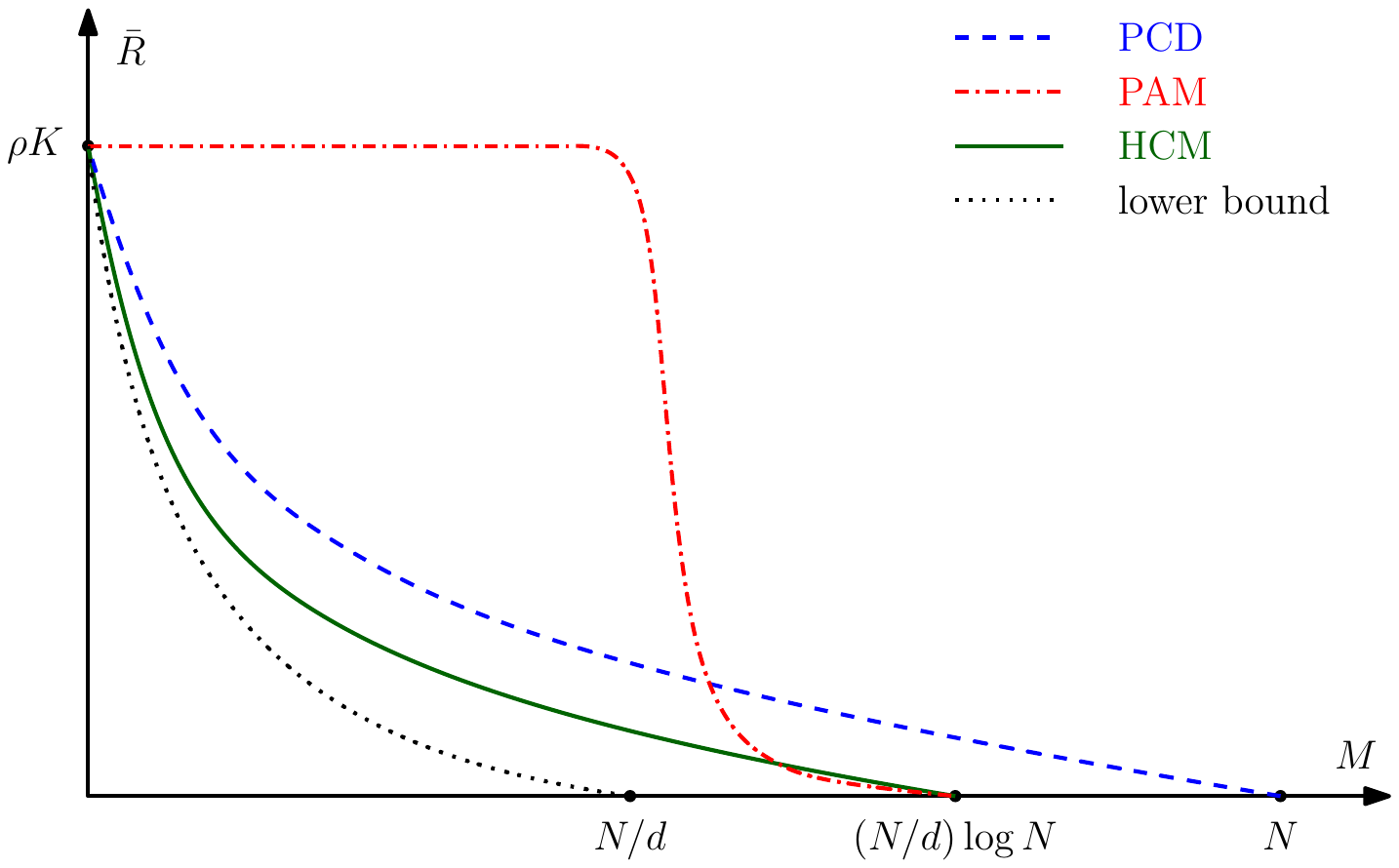}
\caption{Rates achieved by PCD, PAM, and HCM when $\beta\in[0,1)$, along with information-theoretic lower bounds.
HCM is a hybrid scheme described in Section~\ref{sec:hybrid}, and the lower bounds are presented in \appref{app:converse}.
This plot is not numerically generated but is drawn approximately for illustration purposes.}
\label{fig:rate-01}
\end{figure}

\begin{figure}
\centering
\includegraphics[scale=\myfigscale]{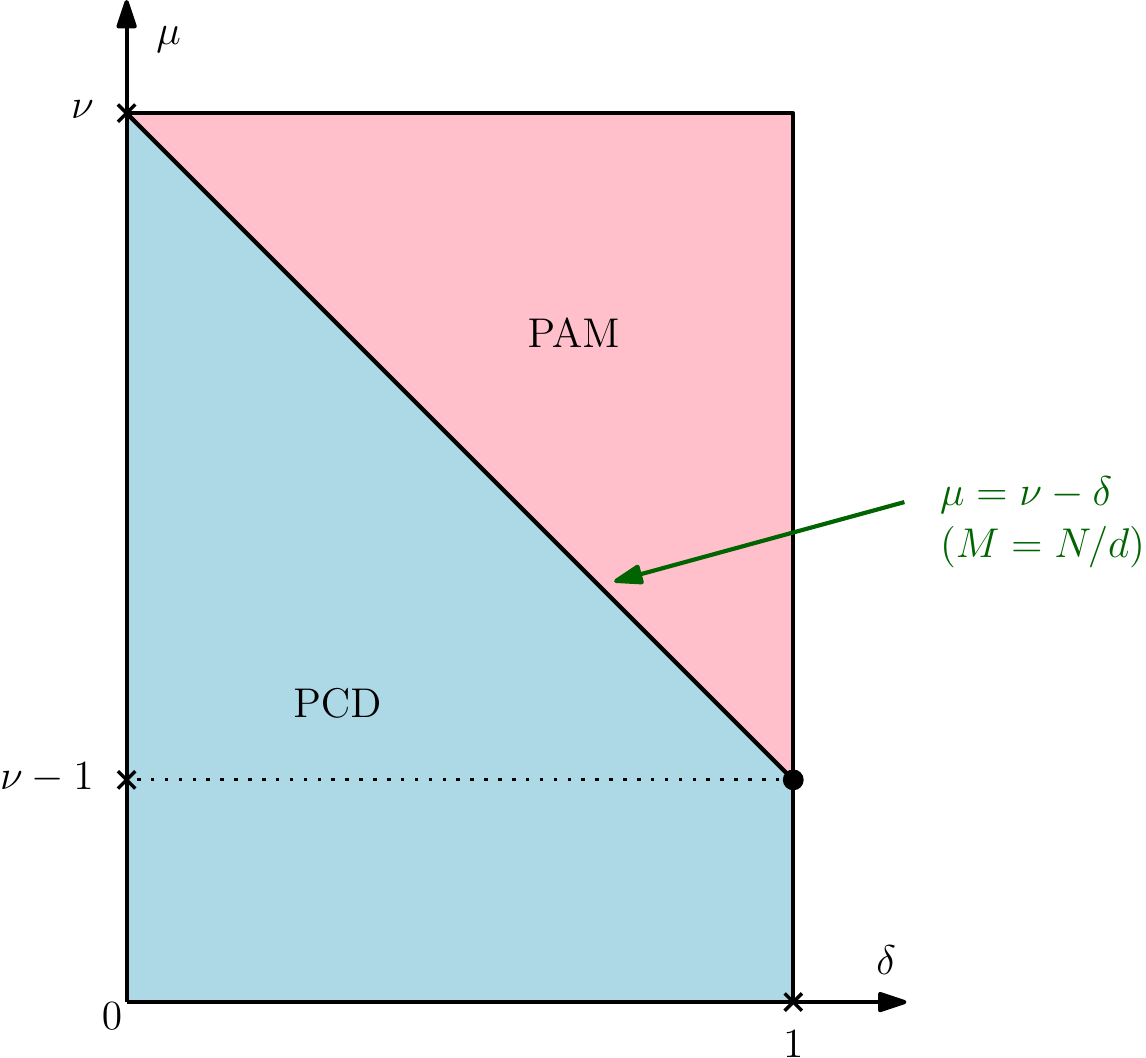}
\caption{The scheme among PCD and PAM that performs better than the other when $\beta\in[0,1)$, in terms of polynomial scaling in $K$.
Here $N=K^\nu$, $d=K^\delta$, and $M=K^\mu$.
Notice that the values in the figure are all independent of $\beta$; the behavior is therefore the same for all $\beta<1$.}
\label{fig:dva-01}
\end{figure}

The rates of PCD and PAM are illustrated in \figurename~\ref{fig:rate-01} for the $\beta\in[0,1)$ case.
We can see that there is a memory threshold $M_0$, with $M_0=\Omega(N/d)$ and $M_0=O( (N/d)\log N )$, such that PCD performs better than PAM for $M<M_0$ while PAM is superior to PCD for $M>M_0$.
Using a poly-$K$ analysis, we can ignore the $\log N$ term and obtain the following result, illustrated in \figurename~\ref{fig:dva-01}.

\begin{theorem}
\label{thm:regimes-01}
When $\beta\in[0,1)$, and considering only a polynomial scaling of the parameters with $K$, PCD outperforms PAM in the regime
\[
\mu \le \nu - \delta,
\]
while PAM outperforms PCD in the opposite regime, where $N=K^\nu$, $d=K^\delta$, and $M=K^\mu$.
\end{theorem}

Note that in some cases PCD and PAM perform equally well, such as when $\mu=\nu$.
However, these are usually edge cases and most of the regimes in Theorem~\ref{thm:regimes-01} are such that one scheme strictly outperforms the other.

Interestingly, under the poly-$K$ analysis, the memory regime where PAM becomes superior to PCD is the regime where PAM achieves a rate of $o(1)$, for any $d$.

\subsection{Approximate Optimality}

In this section, we compare the achievable rates of PCD and PAM schemes to information-theoretic lower bounds and identify regimes in which PCD or PAM is approximately optimal, where approximate optimality is defined below.

\begin{definition}[Approximate Optimality]
\label{def:approx-optim}
We say that a scheme is \emph{approximately optimal} if it can achieve an expected rate $\bar R$ such that
\[
\bar R \le C \cdot \bar R^* + o(1),
\]
where $\bar R^*$ is the optimal expected rate and $C$ is some constant.
\end{definition}

For $\beta\in[0,1)$, we show the approximate optimality of PCD in the small memory regime and that of PAM in the large memory regime.
When $M>\Omega( (N/d)\log N )$, it follows from Theorem~\ref{thm:pam-01} that $\bar R^\mathrm{PAM}=o(1)$, and thus PAM is trivially approximately optimal.
The following theorem states the approximate optimality of PCD when $M<O(N/d)$.

\begin{theorem}
\label{thm:converse}
When $\beta\in[0,1)$ and $M<(1-e^{-1}/2)N/2d$, and for $N\ge10$, the rate achieved by PCD is within a constant factor of the optimum,
\[
\frac{\bar R^\mathrm{PCD}}{\bar R^*}
\le C
\overset{\Delta}{=} \frac{96}{(1-\beta)\rho(1-e^{-1}/2)^2}.
\]
\end{theorem}
Note that the constant $C$ is independent of $K$, $d$, $N$, and~$M$.

Theorem~\ref{thm:converse} can be proved by first reducing the $\beta\in[0,1)$ case to a uniform-popularities setup, and then deriving cut-set lower bounds on the optimal expected rate.
Proof details are given in \appref{app:converse}.

\subsection{A Hybrid Coding and Matching (HCM) Scheme}
\label{sec:hybrid}

So far, we have seen that the two memory regimes $M<O(N/d)$ and $M>\Omega((N/d)\log N)$ require very different schemes: the former requires PCD and the latter requires PAM.
The PCD scheme prioritizes coded delivery at the expense of losing the benefits of adaptive matching, while the PAM scheme leverages adaptive matching but limits itself to an uncoded delivery.
In this section, we introduce a universal scheme that generalizes ideas from both PCD and PAM.
It is a hybrid scheme that combines the benefits of adaptive matching within clusters with the coded caching gains across clusters.
For this reason, we call this scheme Hybrid Coding and Matching (HCM).

The main idea of HCM is to partition files and caches into colors, and then apply a coded caching scheme \emph{within each color} while performing adaptive matching \emph{across colors}.
More precisely, each color consists of a subset of files as well as a subset of the caches of each cluster.
\addchange{The number of caches assigned to a particular color depends on the total popularity of all the files in that color.}
When a user requests a file, the user is matched to an arbitrary cache in its cluster, as long as the cache has the same color as the requested file.
For each color, a coded transmission is then performed to serve all the matched users requesting a file from said color.
Unmatched users are served directly by the server.
This allows us to take advantage of adaptive matching within each cluster as well as obtain coded caching gains across the clusters.

The rate achieved by HCM is given in the following theorem.
It is illustrated in \figurename~\ref{fig:rate-01} along with the rates of PCD and PAM for comparison.
\begin{theorem}
\label{thm:hcm}
For any $\beta\in[0,1)$, HCM can achieve a rate of
\[
\bar R^\mathrm{HCM} = \begin{cases}
\min\left\{ \rho K, \frac{N}{M} - \chi + \frac{K^{-t}}{\sqrt{2\pi}} \right\} & \text{if $M\le\floor{N/\chi}$;}\\
\frac{K^{-t}}{\sqrt{2\pi}} & \text{if $M\ge\ceil{N/\chi}$,}
\end{cases}
\]
where $\chi=\floor{\alpha d/(2(1+t)\log K)}$, for any $t\in[0,t_0]$.
\end{theorem}

While the expression for $\bar R^\mathrm{HCM}$ given in the theorem is rigorous, we can approximate it here for clarity as
\[
\bar R^\mathrm{HCM}
\approx \min\left\{ \rho K,
\left[ \tfrac{N}{M} - \Theta\left(\tfrac{d}{\log K}\right) \right]^+
+ o(1) \right\}.
\]

The proof of Theorem~\ref{thm:hcm} is given in detail in \appref{app:hybrid}, where we provide a rigorous explanation of the HCM scheme.

We will next compare HCM to PCD and PAM.
Notice from \figurename~\ref{fig:rate-01} that HCM is strictly better than PCD for all memory values.
In fact, there is an additive gap between them of about $d/\log K$ for most memory values, and an arbitrarily large multiplicative gap when $M>(N/d)\log K$ where HCM achieves a rate of $o(1)$.
Consequently, HCM is approximately optimal in the regime where PCD is, namely when $M<N/2d$.

Furthermore, HCM is significantly better than PAM in the $M<N/d$ regime: there is a multiplicative gap of up to about $K/d$ between their rates in that regime.
Moreover, HCM achieves a rate of $o(1)$ when $M>(N/d)\log K$.
It is thus trivially approximately optimal in that regime, which includes the regime where PAM is.

\section{The Steep Zipf Case ($\beta>1$)}
\label{sec:steep}

\subsection{Comparing PCD and PAM when $\beta>1$}

\begin{figure}
\centering
\includegraphics[width=\myplotwidth]{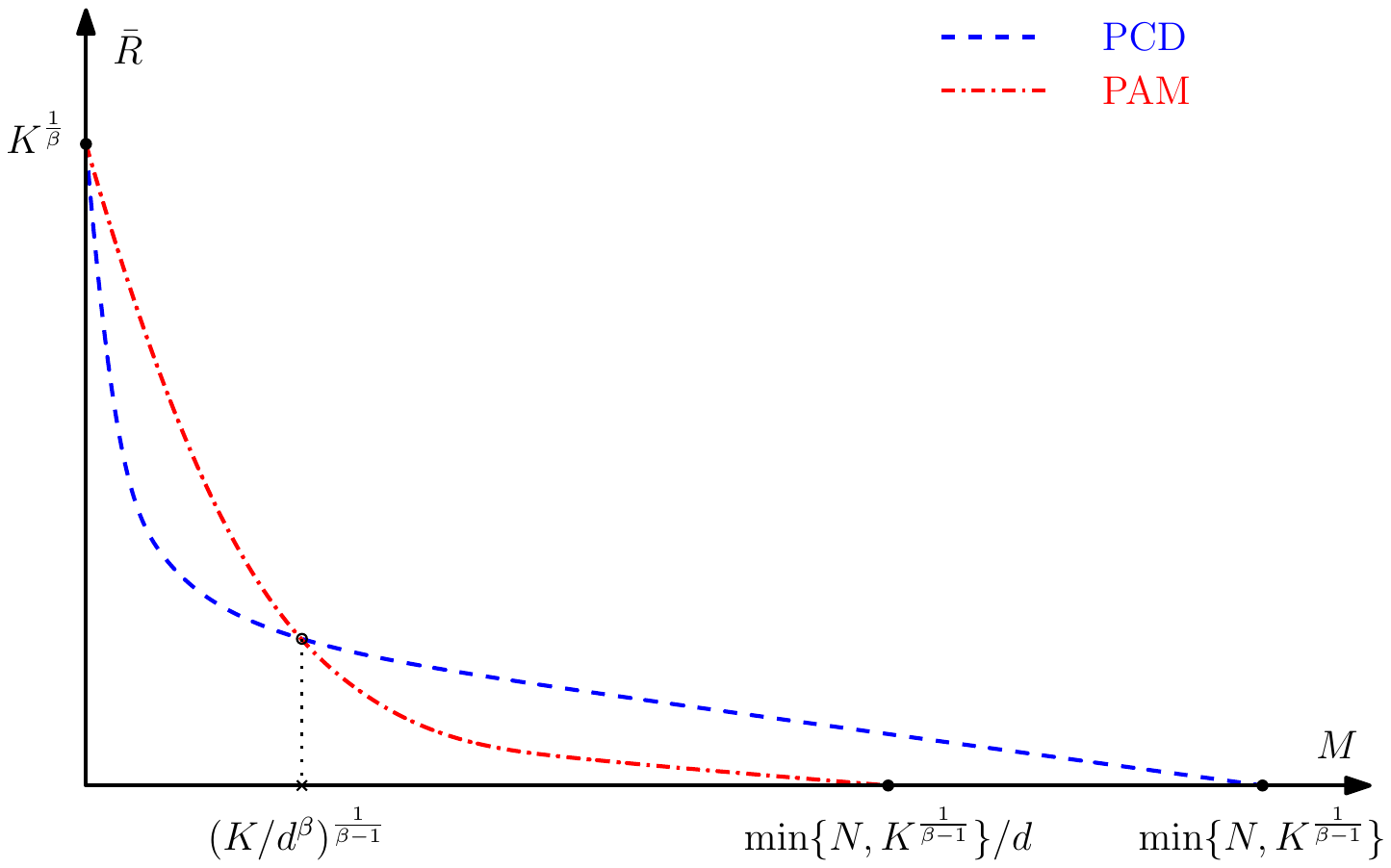}
\caption{Rates achieved by PCD and PAM in the $\beta>1$ case.
Again, this plot is not numerically generated but is drawn approximately for illustration purposes.}
\label{fig:rate-12}
\end{figure}

When $\beta>1$, we restrict ourselves to the case where $d$ is some polynomial in $K$ for convenience.
The following theorems give the rates achieved by PCD and PAM, illustrated in \figurename~\ref{fig:rate-12}.

\begin{theorem}
\label{thm:pcd-12}
When $\beta>1$, the PCD scheme can achieve an expected rate of
\[
\bar R^\mathrm{PCD} = \begin{cases}
K^{1/\beta} & \text{if $0\le M<1$;}\\
\left[ \frac{(KM)^{1/\beta}}{M} - 1 \right]^+ + \frac{K^{-t_0}}{\sqrt{2\pi}} & \text{if $1\le M<N^\beta/K$;}\\
\left[ \frac{N}{M} - 1 \right]^+ + \frac{K^{-t_0}}{\sqrt{2\pi}} & \text{if $M\ge N^\beta/K$.}
\end{cases}
\]
\end{theorem}

Much like Theorem~\ref{thm:pcd-01}, Theorem~\ref{thm:pcd-12} follows from directly applying the coded caching strategy from \cite{Zcodedcaching,ZhangArbitrary}.
Again, the $K^{-t_0}$ term represents the expected number of unmatched users, derived in \apprefthm{app:unmatched}{Lemma}{lemma:pcd-unmatched}.

\begin{theorem}
\label{thm:pam-12}
When $\beta>1$, the PAM scheme can achieve an expected rate of
\[
\bar R^\mathrm{PAM} = \begin{cases}
O\left( \min\left\{ \frac{K}{(dM)^{\beta-1}} , K^{\frac1\beta} \right\} \right) & \text{if $M=O(N/d)$;}\\
o(1) & \text{if $M=\Omega\left( \frac{N\log N}{d} \right)$.}
\end{cases}
\]
\end{theorem}

The proof of Theorem~\ref{thm:pam-12}, given in \appref{app:pam-12}, follows along the same lines as \cite{moharir2016} and involves a generalization from $d=K$ to $d=K^\delta$ for any $0<\delta\le1$.
The idea is to replicate the files across the caches in the cluster, placing more copies for the more popular files, and match the users accordingly.

\begin{figure}
\centering
\includegraphics[scale=\myfigscale]{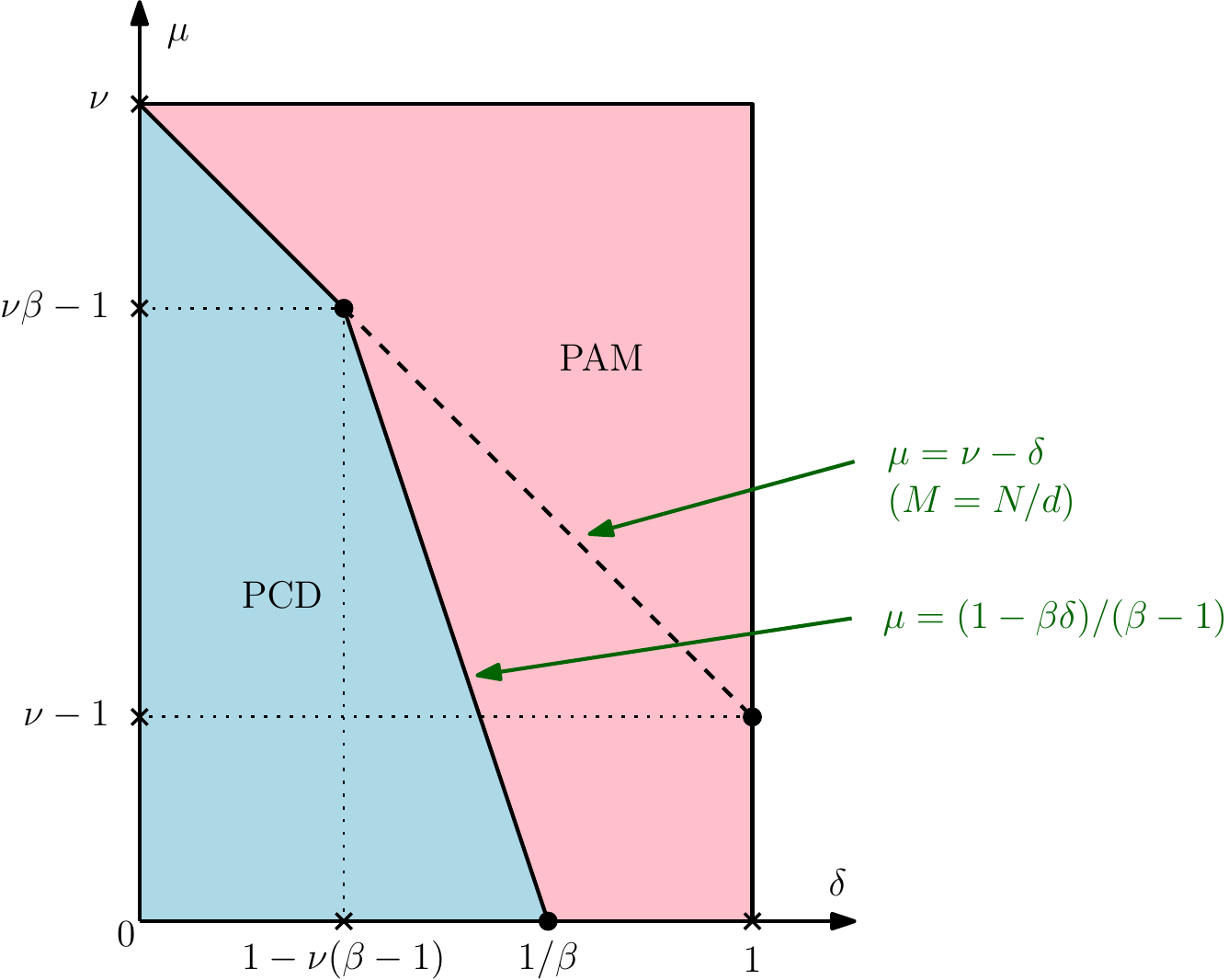}
\caption{The scheme among PCD and PAM that performs better than the other when $\beta>1$ and $\nu<1/(\beta-1)$, in terms of polynomial scaling in $K$.
Here $N=K^\nu$, $d=K^\delta$, and $M=K^\mu$.}
\label{fig:dva-g1}
\end{figure}

As with the $\beta\in[0,1)$ case, we notice that PCD is the better choice when $M$ is small, while PAM is the better choice when $M$ is large.
In fact, by comparing the rate expressions in Theorems~\ref{thm:pcd-12} and~\ref{thm:pam-12} using a poly-$K$ analysis, we obtain the following theorem describing the regimes for which either of PCD or PAM is superior to the other.
The theorem is illustrated in \figurename~\ref{fig:dva-g1} and proved at the end of this subsection.

\begin{theorem}
\label{thm:regimes-12}
When $\beta>1$, and considering only a polynomial scaling of the parameters with $K$, PCD outperforms PAM in the regime
\[
\mu \le \min\left\{ \nu - \delta , (1-\beta\delta)/(\beta-1) \right\},
\]
while PAM outperforms PCD in the opposite regime, where $N=K^\nu$, $d=K^\delta$, and $M=K^\mu$.
\end{theorem}

When comparing Theorems~\ref{thm:regimes-01} and~\ref{thm:regimes-12}, we notice that the case $\beta>1$ has the added constraint $\mu<(1-\beta\delta)/(\beta-1)$ for the regime where PCD is superior to PAM, indicating that there are values of $d$ for which PAM is better than PCD for a larger memory regime under $\beta>1$ as compared to $\beta\in[0,1)$.
This is represented in \figurename~\ref{fig:dva-g1} by the additional line segment joining points $(1-\nu(\beta-1), \nu\beta-1)$ and $(1/\beta,0)$.
As $\beta$ approaches one from above, this line segment tends toward the segment joining points $(1,\nu-1)$ and $(1,0)$.
With it, the regime in which PCD is better than PAM grows until it becomes exactly the regime shown in \figurename~\ref{fig:dva-01} for $\beta\in[0,1)$.
In other words, when $\beta>1$ and as $\beta\to1^+$, the regimes in which PCD or PAM are respectively the better choice become the same regimes as in the $\beta\in[0,1)$ case.
This seemingly continuous transition suggests that, when $\beta=1$, the system should behave similarly to $\beta\in[0,1)$, i.e., \figurename~\ref{fig:dva-01}, at least under a poly-$K$ analysis.

\begin{IEEEproof}[Proof of Theorem~\ref{thm:regimes-12}]
Recall that we are only focusing on a poly-$K$ analysis.
We will define $\sigma^\mathrm{PCD}$ and $\sigma^\mathrm{PAM}$ to be the exponents of $K$ in $\bar R^\mathrm{PCD}$ and $\bar R^\mathrm{PAM}$, respectively, i.e., $\bar R^\mathrm{PCD}=\Theta(K^{\sigma^\mathrm{PCD}})$ and similarly for PAM.
Our goal is to compare $\sigma^\mathrm{PCD}$ to $\sigma^\mathrm{PAM}$.
We can break the proof down into two main cases plus one trivial case.
It can help the reader to follow these cases in \figurename~\ref{fig:dva-g1}.

The trivial case is when the total cluster memory $dM$ is large, specifically $\mu+\delta>\min\{\nu,1/(\beta-1)\}$.
From Theorem~\ref{thm:pam-12}, the PAM rate is then $o(1)$, hence $\sigma^\mathrm{PAM}=0$.
Therefore, PCD cannot perform better than PAM in this case.

In what follows, we assume $\mu+\delta<\min\{\nu,1/(\beta-1)\}$.
We can write the exponents of the rates of PCD and PAM as
\begin{IEEEeqnarray*}{rCl}
\sigma^\mathrm{PCD} &=& \min\left\{ [1-(\beta-1)\mu]/\beta \,,\, \nu - \mu \right\};\\
\sigma^\mathrm{PAM} &=& \min\left\{ 1/\beta \,,\, 1 - (\beta-1)(\delta+\mu) \right\}.
\end{IEEEeqnarray*}
Notice that we always have $\sigma^\mathrm{PCD} \le [1-(\beta-1)\mu]/\beta \le 1/\beta$, and hence it is sufficient to compare $\sigma^\mathrm{PCD}$ to the second term in the minimization in $\sigma^\mathrm{PAM}$.
In other words,
\[
\sigma^\mathrm{PCD} < \sigma^\mathrm{PAM}
\iff
\sigma^\mathrm{PCD} < 1-(\beta-1)(\delta+\mu).
\]
Furthermore, we can write $\sigma^\mathrm{PCD}$ as $\sigma^\mathrm{PCD} = [1-(\beta-1)\mu]/\beta$ if $\mu\le\nu\beta-1$ and $\sigma^\mathrm{PCD} = \nu-\mu$ if $\mu\ge\nu\beta-1$.
For this reason, we split the analysis into a small and a large memory regimes, with the threshold $\mu \lessgtr \nu\beta-1$.

\paragraph*{Large memory: $\mu>\nu\beta-1$}
This case is only possible when $\nu<1/(\beta-1)$ because we always have $\mu\le\nu$.
Here, PCD achieves $\sigma^\mathrm{PCD}=\nu-\mu$.
The constraints on $\mu$ imply:
\begin{IEEEeqnarray*}{rClCl}
\mu &<& \nu-\delta
&\implies&
1 - (\beta-1)(\delta+\mu) > 1-\nu(\beta-1);\\
\mu &>& \nu\beta-1
&\implies&
\nu-\mu < \nu - (\nu\beta-1) = 1 - \nu(\beta-1).
\end{IEEEeqnarray*}
Combining the two inequalities yields
\[
1 - (\beta-1)(\delta+\mu) > 1 - \nu(\beta-1) > \nu-\mu,
\]
and hence $\sigma^\mathrm{PCD}<\sigma^\mathrm{PAM}$.

\paragraph*{Small memory: $\mu<\nu\beta-1$}
In this case, PCD always achieves $\sigma^\mathrm{PCD} = [1-(\beta-1)\mu]/\beta$.
Using some basic algebra, we can show that
\(
\left[1-(\beta-1)\mu\right]/\beta < 1-(\beta-1)(\delta+\mu)
\),
i.e., $\sigma^\mathrm{PCD}<\sigma^\mathrm{PAM}$, if and only if $\mu<(1-\beta\delta)/(\beta-1)$.
\end{IEEEproof}


\subsection{Approximate Optimality}

When $\beta>1$ Theorem~\ref{thm:pam-12} states that $\bar R^\mathrm{PAM}=o(1)$ is achieved for $M>\Omega( \min\{ N\log N, K^{\frac{1}{\beta-1}} \} / d )$.
Thus PAM is trivially approximately optimal in that regime per Definition~\ref{def:approx-optim}.

When the memory is smaller, proving approximate optimality of PCD is more difficult than in the $\beta\in[0,1)$ case.
Indeed, applying a similar technique here to the shallow-Zipf case is insufficient.
This is partly because the number of distinct requested files is close to the number of caches $K$ when $\beta<1$, but becomes much smaller than $K$ when $\beta>1$.

\section{\addchange{Conclusion}}
\label{sec:conclusion}
\addchange{We have studied the caching problem in a new setup in which each user can be matched to one cache in a \emph{cluster} of caches before the delivery phase.
This clustering model bridges two extremes studied in the literature: one in which coded delivery is approximately optimal and one in which adaptive matching with uncoded delivery is approximately optimal.
Our results bring out key insights into the problem.

We observe that there is a natural threshold for when coded delivery is better than adaptive matching, and vice versa.
This threshold depends not only on the cluster size, but also on the content popularity distribution.
It is also fundamental in at least some cases, as evidenced by our approximate optimality results.
Moreover, the existence of a hybrid scheme shows that there are natural ways to combine coded delivery with adaptive matching.}


\appendices

\section{Expected Number of Unmatched Users}
\label{app:unmatched}

In this appendix, we will derive upper bounds on the expected number of unmatched users when using PCD, stated in Lemma~\ref{lemma:pcd-unmatched} below.
The proof of this lemma requires Lemma~\ref{lemma:unmatched}, also stated below, which gives a more general result on the number of unmatched users.

\begin{lemma}
\label{lemma:pcd-unmatched}
When using PCD, the expected number of unmatched users is no greater than $K^{-t_0}/\sqrt{2\pi}$.
\end{lemma}

\begin{lemma}
\label{lemma:unmatched}
If $Y\sim\mathrm{Poisson}(\gamma m)$ users must be matched with $m\ge1$ caches, where $\gamma\in(0,1)$, then the expected number of unmatched users $U=[Y-m]^+$ is bounded by
\[
\mathbb{E}[U] \le \frac{1}{\sqrt{2\pi}} \cdot m \cdot \left( \gamma e^{1-\gamma} \right)^m.
\]
\end{lemma}

Lemma~\ref{lemma:unmatched} is proved in \apprefextra.

\begin{IEEEproof}[Proof of Lemma~\ref{lemma:pcd-unmatched}]
In PCD, at each cluster $c$ we are attempting to match a number of users $Y(c)\sim\mathrm{Poisson}(\rho d)$ to exactly $d$ caches.
Let $U(c)$ denote the number of unmatched users at cluster $c$, and let $U^0=\sum_c U(c)$ be the total number of unmatched users.
The matching in PCD is arbitrary, and so any user can be matched to any cache.
Consequently, $U(c)=[Y(c)-d]^+$ and we can apply Lemma~\ref{lemma:unmatched} directly to obtain
\ifdefined\isdraft
\begin{equation}
\label{eq:unmatched-bound1}
\mathbb{E}[U^0]
= \sum_{c=1}^{K/d} \mathbb{E}[U(c)]
\le \frac{K}{d} \cdot \frac{1}{\sqrt{2\pi}} \cdot d \cdot \left( \rho e^{1-\rho} \right)^d
= \frac{1}{\sqrt{2\pi}} \exp\left\{ \log K + d \log\left( \rho e^{1-\rho} \right) \right\}.
\end{equation}
\else
\begin{IEEEeqnarray*}{rCl}
\mathbb{E}[U^0]
&=& \sum_{c=1}^{K/d} \mathbb{E}[U(c)]
\le \frac{K}{d} \cdot \frac{1}{\sqrt{2\pi}} \cdot d \cdot \left( \rho e^{1-\rho} \right)^d\\
&=& \frac{1}{\sqrt{2\pi}} \exp\left\{ \log K + d \log\left( \rho e^{1-\rho} \right) \right\}.
\IEEEyesnumber\label{eq:unmatched-bound1}
\end{IEEEeqnarray*}
\fi
Note that the function $x\mapsto xe^{1-x}$ is strictly increasing for $x\in(0,1)$.
Since $\rho\in(0,1/2)$, we thus get
\[
\log\left( \rho e^{1-\rho} \right) < \log\left( 2\rho e^{1-2\rho} \right) = -\alpha
< \log(1) = 0.
\]
Applying this to \eqref{eq:unmatched-bound1}, we obtain
\ifdefined\isdraft
\[
\mathbb{E}[U^0]
\le \frac{1}{\sqrt{2\pi}} \exp\left\{ \log K + d\log\left(2\rho e^{1-2\rho}\right) \right\}
\overset{(a)}{\le} \frac{1}{\sqrt{2\pi}} \exp\left\{ \log K - (1+t_0)\log K \right\}
= \frac{1}{\sqrt{2\pi}} K^{-t_0},
\]
\else
\begin{IEEEeqnarray*}{rCl}
\mathbb{E}[U^0]
&\le& \frac{1}{\sqrt{2\pi}} \exp\left\{ \log K + d\log\left(2\rho e^{1-2\rho}\right) \right\}\\
&\overset{(a)}{\le}& \frac{1}{\sqrt{2\pi}} \exp\left\{ \log K - (1+t_0)\log K \right\}
= \frac{1}{\sqrt{2\pi}} K^{-t_0},
\end{IEEEeqnarray*}
\fi
where $(a)$ uses \eqref{eq:d-vs-K}.
This concludes the proof.
\end{IEEEproof}

\section{Details of PAM for $\beta\in[0,1)$
(Proof of Theorem~\ref{thm:pam-01})}
\label{app:pam-01}

First, note that it is always possible to unicast from the server to each user the file that it requested.
Since the expected number of users is $\rho K$, we always have $\bar R^\mathrm{PAM}\le\rho K$.

In what follows, we focus on the regime \addchange{$M\ge N/(1-\beta)d$}.
Recall that the number of requests for file $n$ at cluster $c$ is $u_n(c)$, a Poisson variable with parameter $\rho dp_n$.

In the placement phase, we perform a proportional placement.
Specifically, since each cache can store $M$ files and each cluster consists of $d$ caches, we replicate each file $W_n$ on $d_n=p_ndM$ caches per cluster.
\addchange{Note that $d_n\ge1$ because
\[
p_n\ge p_N=N^{-\beta}/A_N\ge (1-\beta)/N,
\]
by Lemma~\ref{lemma:A_N}, and $M\ge N/(1-\beta)d$.}

In the matching phase, \addchange{the goal is to find an (integral) matching of users to caches, such that users are matched to caches that contain their requested file.
Users that cannot be matched must be served directly by the server.
In order to find an integral matching,} we first construct a fractional matching of users to caches, and then show that this implies the existence of an integral matching\addchange{, which can be found using the Hungarian algorithm}.
We construct the fractional matching by dividing each file $W_n$ into $d_n$ equal parts, and then mapping each request for file $W_n$ to $d_n$ requests, one for each of its parts.
Each user now connects to the $d_n$ caches containing file $W_n$ and retrieves one part from each cache.
This leads to a fractional matching where the total data served by a cache $k$ in cluster $c$ is less than one file if
\begin{equation}
\label{eq:pam01-fractional}
\sum_{W_n\in\mathcal{W}_k} \frac{u_n(c)}{d_n} \le 1,
\end{equation}
where $\mathcal{W}_k$ is the set of files stored on cache $k$.
Let $h(x)=x\log x+1-x$ be the Cram\'{e}r transform of a unit Poisson random variable.
Using the Chernoff bound and the arguments used in the proof of \cite[Proposition~1]{leconte2012}, we have that
\begin{equation}
\label{eq:pam01-chernoff}
\Pr\left\{ \sum_{W_n\in\mathcal{W}_k} \frac{u_n(c)}{d_n} > 1 \right\}
\le e^{-zdM/N},
\end{equation}
where $z=(1-\beta)\rho h(1+(1-\rho)/2\rho)>0$.

To find a matching between the set of requests and the caches, we serve all requests for files that are stored on caches for which \eqref{eq:pam01-fractional} is violated via the server.
For the remaining files, there exists a fractional matching between the set of requests and the caches such that each request is allocated only to caches in the corresponding cluster, and the total data served by each cache is not more than one unit.
By the total unimodularity of adjacency matrix, the existence of a fractional matching implies the existence of an integral matching \cite{schrijver2003}.
We use the Hungarian algorithm to find a matching between the remaining requests and the caches in the corresponding cluster.

Let $\tau_n$ be the probability that at least one of the caches storing file $W_n$ does not satisfy \eqref{eq:pam01-fractional}.
By the union bound, it follows that
\ifdefined\isdraft
$\tau_n \le (KM/N)e^{-zdM/N}$.
\else
\[
\tau_n \le \frac{KM}{N} e^{-zdM/N}.
\]
\fi
By definition,
\begin{equation}
\label{eq:pam01-rate-1}
\bar R^\mathrm{PAM}
\le \sum_{n=1}^N \tau_n
\le KMe^{-zdM/N},
\end{equation}
which concludes the proof of the theorem.

While the above was enough to prove the theorem, we will next provide an additional upper bound on the PAM rate, thus obtaining a tighter expression.

Let $G_c$ be the event that the total number of requests at cluster $c$ is less than $d$, and let $G=\bigcap_c G_c$.
Using the Chernoff bound, we have
\[
\mathbb{E}[G] \ge 1 - \frac{K}{d} e^{-zd},
\]
where $z$ is as defined above.
Conditioned on $G_c$, the number of files that need to be fetched from the server to serve all requests in the cluster is at most $d$.
The rest of this proof is conditioned on $G$.

Let $E_c$ be the event that all caches in cluster $c$ satisfy \eqref{eq:pam01-fractional}.
Using \eqref{eq:pam01-chernoff} and the union bound, we have that
\[
\Pr\{E_c|G\} \ge \Pr\{E_c\} \ge 1 - d e^{-zdM/N},
\]
where $z$ is as defined above.
Conditioned on $E_c$, all the requests in cluster $c$ can be served by the caches.
Therefore,
\[
\mathbb{E}[R|G]
\le \sum_{c=1}^{K/d} d \cdot \Pr\left\{ \overline{E_c} \middle| G \right\}
\le Kd e^{-zdM/N},
\]
where $\overline{A}$ denotes the complement of $A$ for any event $A$.

It follows that
\ifdefined\isdraft
\begin{equation}
\label{eq:pam01-rate-2}
\mathbb{E}[R]
= \mathbb{E}[R|G]\Pr\{G\} + \mathbb{E}[R|\overline{G}]\Pr\{\overline{G}\}
\le \mathbb{E}[R|G] + N\Pr\{\overline{G}\}
\le Kde^{-zdM/N} + \frac{NK}{d}e^{-zd}.
\end{equation}
\else
\begin{IEEEeqnarray*}{rCl}
\mathbb{E}[R]
&=& \mathbb{E}[R|G]\Pr\{G\} + \mathbb{E}[R|\overline{G}]\Pr\{\overline{G}\}\\
&\le& \mathbb{E}[R|G] + N\Pr\{\overline{G}\}\\
&\le& Kde^{-zdM/N} + \frac{NK}{d}e^{-zd}.
\IEEEyesnumber\label{eq:pam01-rate-2}
\end{IEEEeqnarray*}
\fi
Using \eqref{eq:pam01-rate-1} and \eqref{eq:pam01-rate-2}, we obtain
\[
\bar R^\mathrm{PAM}
\le \min\left\{
KMe^{-zdM/N} \,,\,
Kde^{-zdM/N} + \frac{NK}{d} e^{-zd}
\right\},
\]
which is a tighter bound on the PAM rate and implies Theorem~\ref{thm:pam-01}.

\section{Approximate Optimality (Proof of Theorem~\ref{thm:converse})}
\label{app:converse}

In this section, we focus on the case $\beta\in[0,1)$ to prove Theorem~\ref{thm:converse}.
The key idea here is to show that this case can be reduced to a uniform-popularities case.
We will therefore first derive lower bounds for the uniform-popularities setup ($\beta=0$), and then use that result to derive bounds for the more general case.

\subsection{The Uniform-Popularities Case ($\beta=0$)}

When $\beta=0$, we have the following lower bounds.

\begin{lemma}
\label{lemma:uniform-converse}
Let $s\in\{1,\ldots,K/d\}$.
If $N\ge10$, we have the following lower bound on $\bar R^*$:
\[
\bar R^* \ge \frac14 \rho sd \left( 1 - \frac{e^{-1}}{2} - \frac{sdM}{N} \right).
\]
\end{lemma}

Before we prove Lemma~\ref{lemma:uniform-converse}, notice that the lower bounds that it gives are very similar to the ones in \cite{maddah-ali2012}.
In fact, by writing the inequality of Lemma~\ref{lemma:uniform-converse} as
\[
\bar R^* \ge \frac{\rho(1-e^{-1}/2)}{4} d
\cdot \max_{s\in[K/d]} s \left( 1 - \frac{sdM}{(1-e^{-1}/2)N} \right),
\]
we can use the same argument as in \cite{maddah-ali2012} to show
\begin{IEEEeqnarray*}{rCl}
\bar R^*
&\ge& \frac{\rho(1-e^{-1}/2)}{4} d \cdot \frac{1}{12} \min\left\{ \frac{(1-e^{-1}/2)N}{dM} - 1 , \frac{K}{d} \right\}\\
&=& \frac{\rho(1-e^{-1}/2)}{48} \min\left\{ \frac{(1-e^{-1}/2)N}{M} - d, K \right\}.
\IEEEyesnumber\label{eq:uniform-converse}
\end{IEEEeqnarray*}

\begin{IEEEproof}[Proof of Lemma~\ref{lemma:uniform-converse}]
First, consider the following hypothetical scenario.
Let there be a single cache of size $M'$, and suppose that a request profile $\mathbf{u}$ is issued from users all connected to this one cache.
Moreover, assume that we allow the designer to set the cache contents \emph{after} the request profile is revealed; thus both the placement and delivery take place with knowledge of $\mathbf{u}$.
If we send a single message to serve those requests, and denote its rate by $R'(M',\mathbf{u})$, then a cut-set bound shows that
\begin{equation}
\label{eq:hypothetical-cut-set}
R'(M',\mathbf{u}) + M' \ge \gamma(\mathbf{u}),
\end{equation}
where $\gamma(\mathbf{u})$ is the total number of \emph{distinct} files requested in~$\mathbf{u}$.

Since all users share the same resources, if a file can be decoded by one user then it can be decoded by all.
Thus the number of requests for each file is irrelevant for \eqref{eq:hypothetical-cut-set}, as long as it is non-zero.
Furthermore, since both the placement and the delivery are made after the request profile is revealed, the \emph{identity} of the requested files is irrelevant.
Indeed, if $R'(M',\mathbf{u}_1) > R'(M',\mathbf{u}_2)$ for some $\mathbf{u}_1$ and $\mathbf{u}_2$ such that $\gamma(\mathbf{u}_1)=\gamma(\mathbf{u}_2)$, then a simple relabling of the files in $\mathbf{u}_1$ can make it equivalent to $\mathbf{u}_2$, and thus the same rate can be achieved.
Consequently, \eqref{eq:hypothetical-cut-set} can be rephrased using only the \emph{number} of distinct requested files,
\begin{equation}
\label{eq:cut-set-y}
\widetilde R(M',y) + M' \ge y,
\end{equation}
where $\widetilde R(M',y)$ is the rate required to serve requests for $y$ distinct files from one cache of memory $M'$.
Note that $R'(M',\mathbf{u})=\widetilde R(M',\gamma(\mathbf{u}))$ for all $\mathbf{u}$.
Additionally, it can be seen that $\widetilde R(M',y)$ increases as $y$ increases: if $y_1<y_2$, then we can always add $y_2-y_1$ users to request new files, and thus achieve a rate of $\widetilde R(M',y_1)\le\widetilde R(M',y_2)$.

Let us now get back to our original problem.
For convenience, define $\bar R_y = \mathbb{E}_{\mathbf{u}}[R_{\mathbf{u}} | \gamma(\mathbf{u})=y ]$ for every $y$.
Suppose we choose $s\in\{1,\ldots,K/d\}$ different clusters, and we observe the system over $B$ instances.
Over this period, a certain number of users will connect to these clusters and request files; all other users are ignored.
If we denote the resulting request profiles as $\mathbf{u}_1,\ldots,\mathbf{u}_B$, then the rate required to serve all requests is $\sum_{i=1}^B R_{\mathbf{u}_i}$.

Suppose we relax the problem and allow the users to co-operate.
Suppose also that we allow the placement to take place after all $B$ request profiles are made.
This can only reduce the required rate.
Furthermore, this is now an equivalent problem to the hypothetical scenario described at the beginning of the proof.
Therefore, if we denote by $\bar{\mathbf{u}}^B$ the request profile cumulating $\mathbf{u}_1,\ldots,\mathbf{u}_B$, we have
\[
\sum_{i=1}^B R_{\mathbf{u}_i}
\ge R'(sdM, \bar{\mathbf{u}}^B )
= \widetilde R( sdM, \gamma( \bar{\mathbf{u}}^B ) ).
\]
By averaging over the cumulative request profile, we obtain the following bound on any achievable expected rate $\bar R$:
\ifdefined\isdraft
\begin{IEEEeqnarray*}{rCl}
B\bar R
&\ge& \mathbb{E}_{\bar{\mathbf{u}}^B}
\left[ \widetilde R(sdM, \gamma(\bar{\mathbf{u}}^B)) \right]
= \mathbb{E}_{Y_{sB}} \left[ \mathbb{E}_{\bar{\mathbf{u}}^B}
\left[ \widetilde R(sdM, \gamma(\bar{\mathbf{u}}^B)) \middle| \gamma(\bar{\mathbf{u}}^B)=Y_{sB} \right]
\right]\\
&=& \mathbb{E}_{Y_{sB}} \left[ \widetilde R(sdM, Y_{sB}) \right],
\IEEEyesnumber\label{eq:cut-set-exp}
\end{IEEEeqnarray*}
\else
\begin{IEEEeqnarray*}{rCl}
B\bar R
&\ge& \mathbb{E}_{\bar{\mathbf{u}}^B}
\left[ \widetilde R(sdM, \gamma(\bar{\mathbf{u}}^B)) \right]\\
&=& \mathbb{E}_{Y_{sB}} \left[ \mathbb{E}_{\bar{\mathbf{u}}^B}
\left[ \widetilde R(sdM, \gamma(\bar{\mathbf{u}}^B)) \middle| \gamma(\bar{\mathbf{u}}^B)=Y_{sB} \right]
\right]\\
&=& \mathbb{E}_{Y_{sB}} \left[ \widetilde R(sdM, Y_{sB}) \right],
\IEEEyesnumber\label{eq:cut-set-exp}
\end{IEEEeqnarray*}
\fi
where $Y_{sB}$ is a random variable denoting the number of distinct files requested after $B$ instances at $s$ clusters.
Since \eqref{eq:cut-set-exp} holds for all achievable rates $\bar R$, it also holds for $\bar R^*$.

Let us choose $B=\ceil{N/\rho sd}$.
Using Chernoff bounds, we obtain some probabilistic bounds on the number of requested files $Y_{sB}$.
These bounds are given in the following lemma for convenience; the lemma is proved in \apprefextra.
\begin{lemma}
\label{lemma:distinct-files}
If $Y$ denotes the number of distinct requested files by users at $s$ clusters over $B=\ceil{N/\rho sd}$ instances, then, for all $\epsilon>0$,
\[
\Pr\left\{ Y \le (1-e^{-1}-\epsilon)N \right\}
\le e^{-N D(e^{-1}+\epsilon||e^{-1})}.
\]
\end{lemma}

We will now use Lemma~\ref{lemma:distinct-files} to obtain bounds on the expected rate.
Define $\widetilde N = (1-e^{-1}-\epsilon)N$.
From \eqref{eq:cut-set-exp}, we have
\ifdefined\isdraft
\begin{IEEEeqnarray*}{rCl}
\ceil{N/\rho sd} \bar R^*
&\ge& \mathbb{E}_{Y_{sB}}\left[ \widetilde R(sdM, Y_{sB}) \right]
= \sum_{y=1}^N \Pr\{Y_{sB}=y\} \cdot \widetilde R(sdM, y)\\
&\ge& \sum_{y=\widetilde N}^N \Pr\{Y_{sB}=y\} \cdot \widetilde R(sdM, y)
\overset{(a)}{\ge} \widetilde R(sdM, \widetilde N) \sum_{y=\widetilde N}^N \Pr\{Y_{sB}=y\}\\
&\overset{(b)}{\ge}& \widetilde R(sdM, \widetilde N)
\left( 1 - e^{-N D(e^{-1}+\epsilon||e^{-1})} \right)
\overset{(c)}{\ge} (1-o(1)) \left( (1-e^{-1}-\epsilon)N - sdM \right),
\end{IEEEeqnarray*}
\else
\begin{IEEEeqnarray*}{rCl}
\ceil{N/\rho sd} \bar R^*
&\ge& \mathbb{E}_{Y_{sB}}\left[ \widetilde R(sdM, Y_{sB}) \right]\\
&=& \sum_{y=1}^N \Pr\{Y_{sB}=y\} \cdot \widetilde R(sdM, y)\\
&\ge& \sum_{y=\widetilde N}^N \Pr\{Y_{sB}=y\} \cdot \widetilde R(sdM, y)\\
&\overset{(a)}{\ge}& \widetilde R(sdM, \widetilde N) \sum_{y=\widetilde N}^N \Pr\{Y_{sB}=y\}\\
&\overset{(b)}{\ge}& \widetilde R(sdM, \widetilde N)
\left( 1 - e^{-N D(e^{-1}+\epsilon||e^{-1})} \right)\\
&\overset{(c)}{\ge}& (1-o(1)) \left( (1-e^{-1}-\epsilon)N - sdM \right),
\end{IEEEeqnarray*}
\fi
where $(a)$ uses the fact that $\widetilde R(sdM, y)$ can only increase with the number of requested files $y$, $(b)$ follows from Lemma~\ref{lemma:distinct-files}, and $(c)$ is due to \eqref{eq:cut-set-y}.

For a fixed $\epsilon$, the $1-o(1)$ factor approaches $1$ as $N$ grows.
More generally, we can lower-bound it by some constant for a large enough $N$.
For example, if $\epsilon=e^{-1}/2$, then the term is larger than $1/2$ as long as $N\ge10$.
Therefore, we have
\[
\bar R^*
\ge \frac{\left(1-\frac{e^{-1}}{2}\right) N - sdM}{2\ceil{N/\rho sd}}
\ge \frac14 \rho sd \left( 1 - \frac{e^{-1}}{2} - \frac{sdM}{N} \right)
\]
as long as $N\ge10$.
More generally, we can approach
\[
\bar R^*
\ge \frac{\left(1-e^{-1}\right)N - sdM}{\ceil{N/\rho sd}}
\ge \frac12 \rho sd\left( 1 - e^{-1} - \frac{sdM}{N} \right),
\]
if we allow $N$ to be sufficiently large.
\end{IEEEproof}

\subsection{The General Shallow Zipf Case ($\beta\in[0,1)$)}

First, notice that the popularity of each file is
\begin{equation}
\label{eq:uniform}
p_n
\ge p_N
= \frac{N^{-\beta}}{A_N}
\overset{(a)}{\ge} \frac{(1-\beta)N^{-\beta}}{N^{1-\beta}}
= \frac{1-\beta}{N},
\end{equation}
where $A_N$ is defined in Lemma~\ref{lemma:A_N} stated below, and $(a)$ follows from the lemma.

\begin{lemma}
\label{lemma:A_N}
Let $m\ge1$ be an integer and let $\beta\in[0,1)$.
Define $A_m = \sum_{n=1}^m n^{-\beta}$.
Then,
\[
m^{1-\beta} - 1 \le (1-\beta) A_m \le m^{1-\beta}.
\]
\end{lemma}
Lemma~\ref{lemma:A_N} is proved in \apprefextra.

Consider now the following relaxed setup.
Suppose that, for every file $n$, there are $\tilde u_n(c)$ users requesting file $n$ from cluster $c$, where
\[
\tilde u_n(c) \sim \mathrm{Poisson}\left( \frac{(1-\beta)\rho d}{N} \right).
\]
Since $(1-\beta)\rho d/N \le p_n\rho d$ for all $n$ by \eqref{eq:uniform}, the optimal expected rate for this relaxed setup can only be smaller than the rate from the original setup.
Indeed, we can retrieve the original setup by simply creating $\mathrm{Poisson}\left( (p_n-(1-\beta)/N)\rho d \right)$ additional requests for file $n$ at each cluster.

Our relaxed setup is now exactly a uniform-popularities setup except that $\rho$ is replaced by $\rho'=(1-\beta)\rho$, which is still a constant.
Consequently, the information-theoretic lower bounds obtained in Lemma~\ref{lemma:uniform-converse}, and inequality~\eqref{eq:uniform-converse} that follows it, can be directly applied here, giving the following lemma.

\begin{lemma}
\label{lemma:lower-bounds}
When $N\ge10$, the optimal expected rate $\bar R^*$ can be lower-bounded by
\[
\bar R^*
\ge \frac{(1-\beta)\rho(1-e^{-1}/2)}{48}
\min\left\{ \frac{(1-e^{-1}/2)N}{M} - d , K \right\}.
\]
\end{lemma}

When $M<(1-e^{-1}/2)N/2d$, the bound in Lemma~\ref{lemma:lower-bounds} can be further lower-bounded by
\begin{equation}
\label{eq:converse-rstar}
\bar R^*
\ge \frac{(1-\beta)(1-e^{-1}/2)^2\rho}{96}
\cdot \min\left\{ \frac{N}{M}, \rho K \right\}.
\end{equation}
Furthermore, the rate achieved by PCD is upper-bounded by
\begin{equation}
\label{eq:converse-rpcd}
\bar R^\mathrm{PCD}
\le \min\left\{ \rho K, \frac{N}{M} - 1 + \frac{K^{-t_0}}{\sqrt{2\pi}} \right\}
\le \min\left\{ \rho K, \frac{N}{M} \right\}.
\end{equation}
Consequently, combining \eqref{eq:converse-rstar} with \eqref{eq:converse-rpcd} gives us the result of Theorem~\ref{thm:converse}.

\section{Details of HCM (Proof of Theorem~\ref{thm:hcm})}
\label{app:hybrid}

In this section, we are mostly interested in the case where $K$ is larger than some constant.
Specifically, we assume
\begin{equation}
\label{eq:K-lb}
\log K \ge 2g\alpha,
\end{equation}
where $g=(3^{1-\beta}-1)/4^{1-\beta}>0$ is a constant.
In the opposite case, we can achieve a constant rate by simply unicasting to each user the file that it requested.

Let $t\in[0,t_0]$, and let $\chi=\floor{\alpha gd/(2(1+t)\log K)}$.
We will partition the set of files into $\chi$ \emph{colors}.
For each color $x\in\{1,\ldots,\chi\}$, define $\mathcal{W}_x$ as the set of files colored with $x$.
We choose to color the files in an alternating fashion.
More precisely, we choose for each $x\in\{1,\ldots,\chi\}$
\[
\mathcal{W}_x = \left\{ W_n : n \equiv x \pmod{\chi} \right\}.
\]
Notice that $|\mathcal{W}_x|=\floor{N/\chi}$ or $\ceil{N/\chi}$.
We can now define the \emph{popularity of a color $x$} as
\(
\mathcal{P}_x = \sum_{W_n\in\mathcal{W}_x} p_n
\).
The following proposition, proved in \apprefextra, gives a useful lower bound for $\mathcal{P}_x$.
\begin{proposition}
\label{prop:px}
For each $x\in\{1,\ldots,\chi\}$, we have $\mathcal{P}_x \ge g/\chi$.
\end{proposition}

The significance of the above proposition is that the colors will essentially behave as though they are all equally popular.

Next, we partition the caches of each cluster into the same $\chi$ colors.
We choose this coloring in such a way that the number of caches associated with a particular color is proportional to the popularity of that color.
Specifically, exactly $\floor{d\mathcal{P}_x}$ caches in every cluster will be colored with $x$.
This will leave some caches colorless; they are ignored for the entirety of the scheme for analytical convenience.

We can now describe the placement, matching, and delivery phases of HCM.
Consider a particular color $x$.
This color consists of $|\mathcal{W}_x|$ files and $\floor{d\mathcal{P}_x}K/d$ caches in total.
The idea is to perform a Maddah-Ali--Niesen scheme \cite{maddah-ali2012,maddah-ali2013} on each color separately, while matching each user to a cache of the same color of its requested files.
The scheme can be described more formally with the following three steps.

First, in the placement phase, for each color $x$ we perform a Maddah-Ali--Niesen placement of the files $\mathcal{W}_x$ in the caches colored with $x$.

Second, in the matching phase, each user is matched to a cache in its cluster of the same color as the file that the user has requested.
Thus if the user is at cluster $c$ and requests a file from $\mathcal{W}_x$, it is matched to an arbitrary cache from cluster $c$ colored with color $x$.
For each cluster-color pair, if there are more users than caches, then some users must be unmatched.

Third, in the delivery phase, for each color $x$ we perform a Maddah-Ali--Niesen delivery for the users requesting files from $\mathcal{W}_x$.
Next, each unmatched user is served with a dedicated unicast message.
The resulting overall message sent from the server is a concatenation of the messages sent for each color as well as all the unicast messages intended for unmatched users.

Suppose that the broadcast message sent for color $x$ has a rate of $R_x$.
Suppose also that the number of unmatched users is $U^0$.
Then, the total achieved expected rate will be
\begin{equation}
\label{eq:rate-1}
\bar R^\mathrm{HCM}
= \min\left\{
\rho K,
\sum_{x=1}^\chi \mathbb{E}[R_x] + \mathbb{E}[U^0] \right\},
\end{equation}
since $\rho K$ can always be achieved by simply unicasting to every user its requested file.

From \cite{maddah-ali2013}, we know that we can always upper-bound the rate for color $x$ by
\[
R_x \le \left[ \frac{|\mathcal{W}_x|}{M} - 1 \right]^+,
\]
for all $M>0$.
Because $|\mathcal{W}_x|=\floor{N/\chi}$ or $\ceil{N/\chi}$ for all $x$, we obtain
\begin{equation}
\label{eq:rx}
\sum_x R_x \le \begin{cases}
\frac{N}{M} - 1 & \text{if $M\le\floor{N/\chi}$;}\\
0 & \text{if $M\ge\ceil{N/\chi}$;}\\
(N\bmod\chi)\left( \frac{\ceil{N/\chi}}{M} - 1 \right) & \text{otherwise.}
\end{cases}
\end{equation}

All that remains is to find an upper bound for $\mathbb{E}[U^0]$.
Let $Y(c,x)$ represent the number of users at cluster $c$ requesting a file from color $x$.
Since there are $\floor{d\mathcal{P}_x}$ caches at cluster $c$ with color $x$, then exactly $U(c,x)=[Y(c,x)-\floor{d\mathcal{P}_x}]^+$ users will be unmatched.
Thus we can write $U^0$ as
\[
U^0 = \sum_{c=1}^{K/d} \sum_{x=1}^\chi U(c,x).
\]

Before we proceed, it will be helpful to state the following two results, proved in \apprefextra.

\begin{proposition}
\label{prop:poisson-pmf}
If $Y$ is a Poisson variable with parameter $\lambda$, then, for all integers $m\ge1$, the function $\lambda\mapsto\Pr\{Y=m\}$ is increasing in $\lambda$ as long as $\lambda<m$.
\end{proposition}

\begin{proposition}
\label{prop:poisson}
Let $Y$ be a Poisson random variable with parameter $\lambda$, and let $m\ge\lambda$.
Define $U=[Y-m]^+$, i.e., $U=0$ if $Y<m$ and $U=Y-m$ if $Y\ge m$.
Then,
$
\mathbb{E}[U] \le m \Pr\{Y=m\}
$.
\end{proposition}

Notice that $Y(c,x)\sim\mathrm{Poisson}(\rho d \cdot \mathcal{P}_x )$, and that the $Y(c,x)$ users must be matched to $\floor{d\mathcal{P}_x}$.
For convenience, we define $\widetilde Y(c,x) \sim \mathrm{Poisson}(2\rho \cdot \floor{d\mathcal{P}_x})$ and $\widetilde U=[\widetilde Y(c,x)-\floor{d\mathcal{P}_x}]^+$.
Since we have
\[
\rho d \mathcal{P}_x \le 2\rho \floor{d\mathcal{P}_x},
\]
i.e., the Poisson parameter of $\widetilde Y(c,x)$ is at least the Poisson parameter of $Y(c,x)$, then
\ifdefined\isdraft
\[
\mathbb{E}[U(c,x)]
= \sum_{y=\floor{d\mathcal{P}_x}}^\infty y \cdot \Pr\{ Y(c,x) = y \}
\overset{(a)}{\le} \sum_{y=\floor{d\mathcal{P}_x}}^\infty y \cdot \Pr\{ \widetilde Y(c,x) = y \}
= \mathbb{E}[\widetilde U(c,x)],
\]
\else
\begin{IEEEeqnarray*}{rCl}
\mathbb{E}[U(c,x)]
&=& \sum_{y=\floor{d\mathcal{P}_x}}^\infty y \cdot \Pr\{ Y(c,x) = y \}\\
&\overset{(a)}{\le}& \sum_{y=\floor{d\mathcal{P}_x}}^\infty y \cdot \Pr\{ \widetilde Y(c,x) = y \}
= \mathbb{E}[\widetilde U(c,x)],
\end{IEEEeqnarray*}
\fi
where $(a)$ uses Proposition~\ref{prop:poisson-pmf}.

This allows us to apply Proposition~\ref{prop:poisson} on $\widetilde Y(c,x)$ and $\widetilde U(c,x)$ in order to upper-bound the expectation of $U(c,x)$ by
\[
\mathbb{E}[U(c,x)]
\le \mathbb{E}[\widetilde U(c,x)]
\le \frac{1}{\sqrt{2\pi}} \cdot \floor{d\mathcal{P}_x}
\cdot \left( 2\rho e^{1-2\rho} \right)^{\floor{d\mathcal{P}_x}}.
\]
Consequently, we get the upper bound on $\mathbb{E}[U^0]$,
\ifdefined\isdraft
\begin{IEEEeqnarray*}{rCl}
\mathbb{E}[U^0]
&=& \sum_{c=1}^{K/d} \sum_{x=1}^\chi \mathbb{E}[U(c,x)]
\le \sum_{c=1}^{K/d} \sum_{x=1}^\chi
	\frac{1}{\sqrt{2\pi}} \floor{d\mathcal{P}_x}
	\left( 2\rho e^{1-2\rho} \right)^{\floor{d\mathcal{P}_x}}\\
&=& \frac{1}{\sqrt{2\pi}} \sum_{x=1}^\chi
	\frac{K}{d} \cdot \floor{d\mathcal{P}_x}
	\left( 2\rho e^{1-2\rho} \right)^{\floor{d\mathcal{P}_x}}
\le \frac{1}{\sqrt{2\pi}} \sum_{x=1}^\chi
	\mathcal{P}_x \cdot K
	\left( 2\rho e^{1-2\rho} \right)^{\floor{d\mathcal{P}_x}}.
\end{IEEEeqnarray*}
\else
\begin{IEEEeqnarray*}{rCl}
\mathbb{E}[U^0]
&=& \sum_{c=1}^{K/d} \sum_{x=1}^\chi \mathbb{E}[U(c,x)]\\
&\le& \sum_{c=1}^{K/d} \sum_{x=1}^\chi
	\frac{1}{\sqrt{2\pi}} \floor{d\mathcal{P}_x}
	\left( 2\rho e^{1-2\rho} \right)^{\floor{d\mathcal{P}_x}}\\
&=& \frac{1}{\sqrt{2\pi}} \sum_{x=1}^\chi
	\frac{K}{d} \cdot \floor{d\mathcal{P}_x}
	\left( 2\rho e^{1-2\rho} \right)^{\floor{d\mathcal{P}_x}}\\
&\le& \frac{1}{\sqrt{2\pi}} \sum_{x=1}^\chi
	\mathcal{P}_x \cdot K
	\left( 2\rho e^{1-2\rho} \right)^{\floor{d\mathcal{P}_x}}.
\end{IEEEeqnarray*}
\fi
Isolating part of the term in the sum,
\ifdefined\isdraft
\begin{IEEEeqnarray*}{rCl}
K\left(2\rho e^{1-2\rho}\right)^{\floor{d\mathcal{P}_x}}
&=& \exp\left\{ \log K + \log\left( 2\rho e^{1-2\rho} \right)\floor{d\mathcal{P}_x} \right\}
= \exp\left\{ \log K - \alpha\floor{d\mathcal{P}_x} \right\}\\
&\le& \exp\left\{ \log K - \frac{\alpha d\mathcal{P}_x}{2} \right\}
\overset{(a)}{\le} \exp\left\{ \log K - \frac{\alpha d g}{2\chi} \right\}\\
&\overset{(b)}{\le}& \exp\left\{ \log K - \frac{\alpha dg}{2} \cdot \frac{2(1+t)}{\alpha dg}\log K \right\}
= \exp\left\{ \log K - (1+t)\log K \right\}\\
&=& K^{-t},
\end{IEEEeqnarray*}
\else
\begin{IEEEeqnarray*}{rCl}
K\left(2\rho e^{1-2\rho}\right)^{\floor{d\mathcal{P}_x}}
&=& \exp\left\{ \log K + \log\left( 2\rho e^{1-2\rho} \right)\floor{d\mathcal{P}_x} \right\}\\
&\le& \exp\left\{ \log K - \frac{\alpha d\mathcal{P}_x}{2} \right\}\\
&\overset{(a)}{\le}& \exp\left\{ \log K - \frac{\alpha d g}{2\chi} \right\}\\
&\overset{(b)}{\le}& \exp\left\{ \log K - \frac{\alpha dg}{2} \cdot \frac{2(1+t)}{\alpha dg}\log K \right\}\\
&=& \exp\left\{ \log K - (1+t)\log K \right\}
= K^{-t},
\end{IEEEeqnarray*}
\fi
where $(a)$ uses Proposition~\ref{prop:px}, and $(b)$ uses the definition of $\chi$ combined with $\floor{y}\le y$.
We obtain the final upper bound on the expected number of unmatched users,
\begin{equation}
\label{eq:u0}
\mathbb{E}[U^0]
\le \frac{1}{\sqrt{2\pi}} \sum_{x=1}^\chi \mathcal{P}_x K^{-t}
= \frac{K^{-t}}{\sqrt{2\pi}}.
\end{equation}

Finally, we combine \eqref{eq:rx} and \eqref{eq:u0} in \eqref{eq:rate-1} to obtain the rate expression in Theorem~\ref{thm:hcm}, thus completing its proof.

\section{\addchange{Details of PAM for $\beta>1$
(Proof of Theorem~\ref{thm:pam-12})}}
\label{app:pam-12}

At a high level, the PAM strategy consists in storing complete files in the caches, replicating the files across different caches, and then matching the users to the cache that contains their requested file.
Users that cannot be matched to a cache containing their file are served directly from the server.

The above describes PAM strategies very generally; there are many possible schemes for placement and matching within this class of strategies.
In this paper, we adopt for $\beta>1$ a strategy that performs a \emph{knapsack storage (KS)} placement phase that is based on the knapsack problem, and a \emph{match least popular (MLP)} matching phase in which matching is done for the least popular files first.
We refer to this PAM scheme as KS+MLP.

\subsection{Placement Phase: Knapsack Storage}

We split the KS policy into two parts.
In the first part, we determine how many copies of each file will be stored per cluster.
In the second part, we determine which caches in each cluster will store each file.

\subsubsection{KS Part 1}
\label{sec:pam-ks-1}

The first part of the knapsack storage policy determines how many caches in each cluster store each file by solving a fractional knapsack problem, described as follows.
\addchange{The idea is to find a fractional matching $(x_1,\ldots,x_N)$, where $x_n\in[0,1]$ denotes the fraction of file $W_n$ that will be stored in some caches in the cluster (the remaining $1-x_n$ fraction is not stored anywhere in the cluster).
File $W_n$ will thus take up a memory $x_n$ in each cache that decides to store it, which will be reflected in the \emph{weight} parameter of the knapsack problem.
On the other hand, we will benefit from having stored a fraction of $W_n$ if it is requested by some users, and this will be reflected in the \emph{value} parameter of the knapsack problem.}

The parameters of the fractional knapsack problem are \addchange{hence} a value $v_n$ and a weight $w_n$ associated with each file $W_n$, defined as follows.

The value $v_n$ of file $W_n$ is the probability that $W_n$ is requested by at least one user in a cluster,
\(
v_n = 1 - \left( 1 - p_n \right)^d
\).

The weight $w_n$ of file $W_n$ represents the number of caches in which $W_n$ will be stored, should the policy decide to store it.
If we decide to store a file, we would like to make sure that all requests for that file can be served by the caches, so that it need not be transmitted by the server.
To ensure this, we fix $w_n$ to be large enough so that, with probability going to one as $K\to\infty$, the number of requests for $W_n$ is no larger than $w_n$.
We thus choose the following values for $w_n$:
\[
w_n = \begin{cases}
d & \text{if $n=1$;}\\
\ceil{\left(1+\frac{p_1}{2}\right)\rho d p_n} & \text{if $2\le n\le N_1$;}\\
\ceil{4p_1(\log d)^2} & \text{if $N_1 < n \le N_2$;}\\
1 & \text{if $N_2 < n \le N$,}
\end{cases}
\]
where $N_1$ and $N_2$ are defined as
\begin{equation}
\label{eq:pam-thresholds}
N_1 = \frac{d^{1/\beta}}{p_1(\log d)^{2/\beta}};\quad
N_2 = d^{(1+1/\beta)/2}.
\end{equation}

Using the above parameters $v_n$ and $w_n$, we solve the following knapsack problem:
\begin{IEEEeqnarray*}{c"l}
\underset{x_1,\ldots,x_N\in[0,1]}{\text{maximize}} & \sum_{n=1}^N v_nx_n\\
\text{subject to}
& \sum_{n=1}^N w_nx_n \le dM;
\IEEEyesnumber\label{eq:knapsack}
\end{IEEEeqnarray*}
\addchange{Note that the first inequality in the constraints represents the cluster memory constraint.}
Then, the number of copies of file $W_n$ that will be present in each cluster is $c_n=\floor{x_n}w_n$.
Note that $c_n$ is hence either zero or $w_n$.

\subsubsection{KS Part 2}

The second part of the knapsack storage policy is to determine which caches store each file.
We will focus on one arbitrary cluster, but the same placement is done in each cluster.
To do that, define the multiset $\mathcal{S}$ containing exactly $c_n$ copies of each file index $n$.
Let us order the elements of $\mathcal{S}$ in increasing order, and call the resulting ordered list $(n_1,\ldots,n_{\floor{dM}})$.
Then, for each $r$, we store file $W_{n_r}$ in cache $( (r-1)\bmod d)+1$ of the cluster.

\subsection{Matching and Delivery Phases: Match Least Popular}

In the matching phase, we use the Match Least Popular (MLP) policy, the key idea of which is to match users to caches starting with the users requesting the least popular files.
Algorithm~\ref{alg:pam-mlp} gives the precise description of MLP.

\begin{algorithm}[t]
\caption{The Match Least Popular (MLP) matching policy for a fixed cluster $c$.}
\label{alg:pam-mlp}
\begin{algorithmic}[1]
\Require Number of requests $u_n(c)$ for file $W_n$, for each $n$, at the cluster
\Ensure Matching of users to caches
\State Set $\mathcal{K}_n\subseteq\{1,\ldots,d\}$ to be the set of caches containing file $W_n$, for each $n$
\State \emph{Loop over all files from least to most popular:}
\For{$n\gets N,N-1,\ldots,1$}
\State \emph{Loop over all requests for file $n$:}
\For{$v\gets 1,\ldots,u_n(c)$}
\If{$\mathcal{K}_n\not=\emptyset$} \label{line:available-cache}
\State Pick $k\in\mathcal{K}_n$ uniformly at random
\State Match a user requesting file $W_n$ to cache $k$
\State \emph{Cache $k$ is no longer available:}
\ForAll{$n'\in\{1,\ldots,N\}$}
\State $\mathcal{K}_{n'}\gets\mathcal{K}_{n'}\setminus\{k\}$
\EndFor
\EndIf
\EndFor
\EndFor
\end{algorithmic}
\end{algorithm}

At the end of Algorithm~\ref{alg:pam-mlp}, some users will be unmatched, particularly those for which the condition on line~\ref{line:available-cache} fails.
Any file requested by an unmatched user will be broadcast directly from the server.

\subsection{Expected Rate Achieved by KS+MLP}

\begin{lemma}
\label{lemma:chernoff}
Let $X$ be a Poisson random variable with mean $\mu$, and let $\epsilon\in(0,1)$ be arbitrary.
Then,
\[
\Pr\left\{ X \ge (1+\epsilon)\mu \right\} \le e^{-\mu h(1+\epsilon)},
\]
where $h(x)=x\log x + 1 - x$.
\end{lemma}
\begin{IEEEproof}
The lemma follows from the Chernoff bound.
\end{IEEEproof}

\begin{lemma}
\label{lemma:pam-n-requests}
Recall that $u_n(c)$ denotes the number of users requesting file $W_n$ from cluster $c$.
Consider an arbitrary cluster $c$.
Let $E_1$ denote the event that
\begin{IEEEeqnarray*}{rCl'l}
u_n(c) &\le& \left( 1 + \frac{p_1}{4} \right)dp_n & \text{for all } 1\le n\le N_1; \text{ and}\\
u_n(c) &\le& 2p_1(\log d)^2 & \text{for all } N_1<n\le N_2,
\end{IEEEeqnarray*}
where $N_1$ and $N_2$ are as defined in \eqref{eq:pam-thresholds}.
Then,
\[
\Pr\left\{ E_1 \right\} = 1 - N e^{-\Omega( (\log d)^2 )}.
\]
\end{lemma}

\begin{lemma}
\label{lemma:mlp-prob}
Let $\mathcal{R}=\{n : x_n=1\}$, where $x_n$ is the solution to the fractional knapsack problem solved in Appendix~\ref{sec:pam-ks-1}.
Let $E_2$ denote the event that, in a given cluster, the MLP policy matches \emph{all} requests for all files in $\mathcal{R}$ to caches.
Then,
\[
\Pr\{E_2\} = 1 - Ne^{-\Omega( (\log d)^2 )}.
\]
\end{lemma}
Lemma~\ref{lemma:pam-n-requests} and~\ref{lemma:mlp-prob} are proved in \apprefextra.

From Lemma~\ref{lemma:mlp-prob}, we know that, for $d$ large enough, with probability at least $1-Ne^{-\Omega( (\log d)^2 )}$, in a given cluster all requests for the files cached by the KS+MLP policy are matched to caches.
Let $\widetilde N$ be the number of files not in $\mathcal{R}$ (i.e., that are not cached) that are requested at least once.
By the union bound over the $K/d$ clusters,
\ifdefined\isdraft
\[
\bar R^\mathrm{PAM}
\le \mathbb{E}[\widetilde N] + \frac{K^2}{d}\left( 1 - \Pr\{E_2\} \right)
\le \mathbb{E}[\widetilde N] + \frac{NK^2}{d}e^{-\Omega( (\log d)^2 )}.
\]
\else
\begin{IEEEeqnarray*}{rCl}
\bar R^\mathrm{PAM}
&\le& \mathbb{E}[\widetilde N] + \frac{K^2}{d}\left( 1 - \Pr\{E_2\} \right)\\
&\le& \mathbb{E}[\widetilde N] + \frac{NK^2}{d}e^{-\Omega( (\log d)^2 )}.
\end{IEEEeqnarray*}
\fi

After solving the fractional knapsack problem, defined in \eqref{eq:knapsack}, as a function of $N$, $K$, $d$, $\beta$, and $M$, we can determine the set $\mathcal{R}$.
For a given $\mathcal{R}$, we then have
\[
\mathbb{E}[\widetilde N]
= \sum_{n\notin\mathcal{R}} \left( 1 - (1-p_n)^K \right).
\]
We hence obtain the following bound on the expected rate:
\[
\bar R^\mathrm{PAM}
\le \sum_{n\notin\mathcal{R}} \left( 1 - \left( 1-\frac{n^{-\beta}}{A_N} \right)^K \right)
+ \frac{NK^2}{d} e^{-\Omega( (\log d)^2 )}.
\]
When $N$ and $d$ are polynomial in $K$, then the second term is $o(1)$, and solving the fractional knapsack problem yields the result of Theorem~\ref{thm:pam-12}.

\ifdefined\isextended
\section{Extra Proofs}
\label{app:extra}

\begin{IEEEproof}[Proof of Lemma~\ref{lemma:unmatched}]
By using Proposition~\ref{prop:poisson} (stated in Appendix~\ref{app:hybrid}) with $\lambda=\gamma m$, we have
\[
\mathbb{E}[U]
\le m \Pr\{ Y = m \}
= m \cdot \frac{ (\gamma m)^m e^{-\gamma m} }{m!}.
\]
Using Stirling's approximation, we have
\[
m! \ge \sqrt{2\pi} m^{m+\frac{1}{2}} e^{-m} \ge \sqrt{2\pi} m^m e^{-m},
\]
which yields
\[
\mathbb{E}[U]
\le m \cdot \frac{ (\gamma m)^m e^{-\gamma m} }{\sqrt{2\pi} m^m e^{-m} }
= \frac{1}{\sqrt{2\pi}} \cdot m \cdot \left( \gamma e^{1-\gamma} \right)^m,
\]
thus concluding the proof.
\end{IEEEproof}

\begin{IEEEproof}[Proof of Proposition~\ref{prop:px}]
Choose any $x\in\{1,\ldots,\chi\}$.
Starting with the definition of $\mathcal{P}_x$, we have
\ifdefined\isdraft
\begin{IEEEeqnarray*}{rCl}
\mathcal{P}_x
&=& \sum_{W_n\in\mathcal{W}_x} p_n
= \frac{1}{A_N} \sum_{k=0}^{|\mathcal{W}_x|-1} \left( k\chi + x \right)^{-\beta}
\ge \frac{1}{A_N} \sum_{k=0}^{|\mathcal{W}_x|-1} \left( k\chi + \chi \right)^{-\beta}
= \frac{\chi^{-\beta}}{A_N} \sum_{k=0}^{|\mathcal{W}_x|-1} \left( k+1 \right)^{-\beta}\\
&=& \chi^{-\beta} \cdot \frac{A_{|\mathcal{W}_x|}}{A_N}
\overset{(a)}{\ge} \chi^{-\beta} \cdot \frac{|\mathcal{W}_x|^{1-\beta}-1}{N^{1-\beta}}
\overset{(b)}{\ge} \chi^{-\beta} \cdot \frac{(N/\chi-1)^{1-\beta}-1}{N^{1-\beta}}\\
&=& \frac{1}{\chi} \cdot \left[ \left( 1 - \frac{\chi}{N} \right)^{1-\beta} - \left( \frac{\chi}{N} \right)^{1-\beta} \right]
\overset{(c)}{\ge} \frac{1}{\chi} \left[ \left( 1 - \frac{g\alpha}{2\log K} \right)^{1-\beta} -\left( \frac{g\alpha}{2\log K} \right)^{1-\beta} \right]\\
&\overset{(d)}{\ge}& \frac{1}{\chi} \left[ \left( 1 - \frac{1}{4} \right)^{1-\beta} - \left( \frac{1}{4} \right)^{1-\beta} \right]
= \frac{g}{\chi},
\end{IEEEeqnarray*}
\else
\begin{IEEEeqnarray*}{rCl}
\mathcal{P}_x
&=& \sum_{W_n\in\mathcal{W}_x} p_n\\
&=& \frac{1}{A_N} \sum_{k=0}^{|\mathcal{W}_x|-1} \left( k\chi + x \right)^{-\beta}\\
&\ge& \frac{1}{A_N} \sum_{k=0}^{|\mathcal{W}_x|-1} \left( k\chi + \chi \right)^{-\beta}\\
&=& \frac{\chi^{-\beta}}{A_N} \sum_{k=0}^{|\mathcal{W}_x|-1} \left( k+1 \right)^{-\beta}\\
&=& \chi^{-\beta} \cdot \frac{A_{|\mathcal{W}_x|}}{A_N}\\
&\overset{(a)}{\ge}& \chi^{-\beta} \cdot \frac{|\mathcal{W}_x|^{1-\beta}-1}{N^{1-\beta}}\\
&\overset{(b)}{\ge}& \chi^{-\beta} \cdot \frac{(N/\chi-1)^{1-\beta}-1}{N^{1-\beta}}\\
&=& \frac{1}{\chi} \cdot \left[ \left( 1 - \frac{\chi}{N} \right)^{1-\beta} - \left( \frac{\chi}{N} \right)^{1-\beta} \right]\\
&\overset{(c)}{\ge}& \frac{1}{\chi} \left[ \left( 1 - \frac{g\alpha}{2\log K} \right)^{1-\beta} -\left( \frac{g\alpha}{2\log K} \right)^{1-\beta} \right]\\
&\overset{(d)}{\ge}& \frac{1}{\chi} \left[ \left( 1 - \frac{1}{4} \right)^{1-\beta} - \left( \frac{1}{4} \right)^{1-\beta} \right]\\
&=& \frac{g}{\chi},
\end{IEEEeqnarray*}
\fi
where $(a)$ uses Lemma~\ref{lemma:A_N}, $(b)$ uses the fact that $|\mathcal{W}_x|\ge\floor{N/\chi}\ge N/\chi-1$ for all $x$, $(c)$ uses the definition of $\chi$ as well as $N\ge d$ and $t\ge0$, and $(d)$ uses \eqref{eq:K-lb}.
\end{IEEEproof}

\begin{IEEEproof}[Proof of Proposition~\ref{prop:poisson-pmf}]
Define $f_m(\lambda)=\Pr\{Y=m\}$ when $Y$ is Poisson with parameter $\lambda$, i.e., $f_m(\lambda)=\lambda^me^{-\lambda}/m!$.
Then,
\begin{IEEEeqnarray*}{rCl}
f_m'(y)
&=& \frac{1}{m!} \left( m\lambda^{m-1}e^{-\lambda} - \lambda^me^{-\lambda} \right)
= \frac{\lambda^{m-1}e^{-\lambda}}{m!} \left( m - \lambda \right).
\end{IEEEeqnarray*}
Consequently, $f_m'(y)>0$ if and only if $\lambda<m$, and hence $\Pr\{Y=m\}$ increases with $\lambda$ as long as $\lambda<m$.
\end{IEEEproof}

\begin{IEEEproof}[Proof of Proposition~\ref{prop:poisson}]
\addchange{%
Before we prove Proposition~\ref{prop:poisson}, we need the following useful result, proved later in the appendix.
\begin{proposition}
\label{prop:poisson-tailE}
If $Y$ is a Poisson variable with parameter $\lambda$, then for all integers $m\ge1$,
\[
\mathbb{E}\left[ Y\middle|Y\ge m \right]
= m\Pr\{Y=m|Y\ge m\} + \lambda.
\]
\end{proposition}
We can now prove Proposition~\ref{prop:poisson}.}

Define $V$ such that $V=0$ if $Y<m$ and $V=1$ if $Y\ge m$.
Using the tower property of expectation,
\ifdefined\isdraft
\begin{IEEEeqnarray*}{rCl}
\mathbb{E}[U]
&=& \mathbb{E}\left[ \mathbb{E}\left[ U \middle| V \right] \right]
= \Pr\{V=0\} \mathbb{E}\left[ U \middle| V=0 \right]
+ \Pr\{V=1\} \mathbb{E}\left[ U \middle| V=1 \right]\\
&\overset{(a)}{=}& 0 + \Pr\{V=1\} \mathbb{E}\left[ Y-m \middle| V=1 \right]
= \Pr\{Y\ge m\} \mathbb{E}\left[ Y - m \middle| Y \ge m \right]\\
&=& \Pr\{ Y\ge m \} \left( \mathbb{E}[Y|Y\ge m] - m \right)
\overset{(b)}{=} \Pr\{ Y\ge m \} \left( m \Pr\{Y=m|Y\ge m\} + \lambda - m \right)\\
&\overset{(c)}{\le}& \Pr\{Y\ge m\} \cdot m \Pr\{Y=m|Y\ge m\}
= m \Pr\{Y=m\},
\end{IEEEeqnarray*}
\else
\begin{IEEEeqnarray*}{rCl}
\mathbb{E}[U]
&=& \mathbb{E}\left[ \mathbb{E}\left[ U \middle| V \right] \right]\\
&=& \Pr\{V=0\} \mathbb{E}\left[ U \middle| V=0 \right]
+ \Pr\{V=1\} \mathbb{E}\left[ U \middle| V=1 \right]\\
&\overset{(a)}{=}& 0 + \Pr\{V=1\} \mathbb{E}\left[ Y-m \middle| V=1 \right]\\
&=& \Pr\{Y\ge m\} \mathbb{E}\left[ Y - m \middle| Y \ge m \right]\\
&=& \Pr\{ Y\ge m \} \left( \mathbb{E}[Y|Y\ge m] - m \right)\\
&\overset{(b)}{=}& \Pr\{ Y\ge m \} \left( m \Pr\{Y=m|Y\ge m\} + \lambda - m \right)\\
&\overset{(c)}{\le}& \Pr\{Y\ge m\} \cdot m \Pr\{Y=m|Y\ge m\}\\
&=& m \Pr\{Y=m\},
\end{IEEEeqnarray*}
\fi
where $(a)$ uses the definition of $U$ given the different values of $V$, $(b)$ uses Proposition~\ref{prop:poisson-tailE}, and $(c)$ uses $\lambda\le m$.
\end{IEEEproof}

\begin{IEEEproof}[Proof of Proposition~\ref{prop:poisson-tailE}]
First notice that
\ifdefined\isdraft
\begin{equation}
\label{eq:poisson-prop}
\lambda \Pr\{Y=m-1\}
= \lambda \cdot \frac{\lambda^{m-1}e^{-\lambda}}{(m-1)!}
= \frac{\lambda^m e^{-\lambda}}{m!} \cdot m
= m \Pr\{Y=m\}.
\end{equation}
\else
\begin{IEEEeqnarray*}{rCl}
\lambda \Pr\{Y=m-1\}
&=& \lambda \cdot \frac{\lambda^{m-1}e^{-\lambda}}{(m-1)!}\\
&=& \frac{\lambda^m e^{-\lambda}}{m!} \cdot m\\
&=& m \Pr\{Y=m\}. \IEEEyesnumber\label{eq:poisson-prop}
\end{IEEEeqnarray*}
\fi
We can now write the conditional expectation as
\ifdefined\isdraft
\begin{IEEEeqnarray*}{rCl}
\mathbb{E}\left[ Y \middle| Y\ge m \right]
&=& \sum_{y=m}^\infty y \cdot \frac{\Pr\{Y=y\}}{\Pr\{Y\ge m\}}
= \frac{1}{\Pr\{Y\ge m\}} \sum_{y=m}^\infty y \cdot \frac{\lambda^y e^{-\lambda}}{y!}\\
&=& \frac{1}{\Pr\{Y\ge m\}} \lambda \sum_{y=m}^\infty \frac{\lambda^{y-1} e^{-\lambda}}{(y-1)!}
= \frac{1}{\Pr\{Y\ge m\}} \lambda \cdot \Pr\{Y\ge m-1\}\\
&=& \frac{\lambda \Pr\{Y=m-1\} + \lambda \Pr\{Y\ge m\}}{\Pr\{Y\ge m\}}
\overset{(a)}{=} \frac{m\Pr\{Y=m\} + \lambda\Pr\{Y\ge m\}}{\Pr\{Y\ge m\}}\\
&=& m\Pr\{Y=m|Y\ge m\} + \lambda,
\end{IEEEeqnarray*}
\else
\begin{IEEEeqnarray*}{rCl}
\mathbb{E}\left[ Y \middle| Y\ge m \right]
&=& \sum_{y=m}^\infty y \cdot \frac{\Pr\{Y=y\}}{\Pr\{Y\ge m\}}\\
&=& \frac{1}{\Pr\{Y\ge m\}} \sum_{y=m}^\infty y \cdot \frac{\lambda^y e^{-\lambda}}{y!}\\
&=& \frac{1}{\Pr\{Y\ge m\}} \lambda \sum_{y=m}^\infty \frac{\lambda^{y-1} e^{-\lambda}}{(y-1)!}\\
&=& \frac{1}{\Pr\{Y\ge m\}} \lambda \cdot \Pr\{Y\ge m-1\}\\
&=& \frac{\lambda \Pr\{Y=m-1\} + \lambda \Pr\{Y\ge m\}}{\Pr\{Y\ge m\}}\\
&\overset{(a)}{=}& \frac{m\Pr\{Y=m\} + \lambda\Pr\{Y\ge m\}}{\Pr\{Y\ge m\}}\\
&=& m\Pr\{Y=m|Y\ge m\} + \lambda,
\end{IEEEeqnarray*}
\fi
where $(a)$ uses \eqref{eq:poisson-prop}
\end{IEEEproof}

\begin{IEEEproof}[Proof of Lemma~\ref{lemma:distinct-files}]
Recall that we are considering $s$ clusters and $B=\ceil{N/\rho sd}$ instances of the problem.
Let $U_n$ denote the number of requests for file $n$ by users in these clusters across the $B$ instances.
We can see that $U_n$ is a Poisson variable with parameter
\[
\lambda = \frac{\rho d}{N} \cdot sB = \frac{\rho sd}{N} \cdot \ceil{\frac{N}{\rho sd}}.
\]
Note that $\lambda\ge1$.

Let $Z_n$ be equal to one if $U_n\ge1$, i.e., if file $n$ was requested at least once, and equal to zero otherwise.
Thus $Z_n$ is a Bernoulli variable with parameter $\Pr\{U_n\ge1\}=1-e^{-\lambda}$.
Then, the total number of distinct requested files can be written as $Y=\sum_n Z_n$.

Let $\epsilon>0$ be arbitrary, and define $\eta=1-e^{-1}-\epsilon$.
We can now use the Chernoff bound to write, for every $t>0$,
\begin{IEEEeqnarray*}{rCl}
\Pr\left\{ Y \le \eta N \right\}
&\le& e^{t\eta N} \cdot \mathbb{E}\left[ e^{-tY} \right]
= e^{t\eta N} \cdot \prod_{n=1}^N \mathbb{E}\left[ e^{-tZ_n} \right].
\end{IEEEeqnarray*}
The expression inside the product is
\begin{IEEEeqnarray*}{rCl}
\mathbb{E}\left[ e^{-tZ_n} \right]
&=& e^{-\lambda} \cdot 1 + (1-e^{-\lambda}) \cdot e^{-t}
= e^{-\lambda} ( 1 - e^{-t} ) + e^{-t}\\
&\le& e^{-1} ( 1 - e^{-t} ) + e^{-t}
= e^{-1} + e^{-t}(1-e^{-1}),
\end{IEEEeqnarray*}
where the inequality is due to $\lambda\ge1$ and the fact that the function $\lambda\mapsto e^{-\lambda}(1-e^{-t})$ decreases as $\lambda$ increases, for all $t>0$.
Consequently,
\begin{IEEEeqnarray*}{rCl}
\Pr\left\{ Y \le \eta N \right\}
&\le& \left[ e^{t\eta} \cdot \left( e^{-1} + e^{-t}(1-e^{-1}) \right) \right]^N.
\end{IEEEeqnarray*}
By choosing $t$ such that
\[
e^t = \frac{e^{-1}+\epsilon}{e^{-1}} \cdot \frac{1-e^{-1}}{1-(e^{-1}+\epsilon)},
\]
we get
\begin{IEEEeqnarray*}{rCl}
\Pr\left\{ Y \le \eta N \right\}
&\le& \left[ e^{-D(e^{-1}+\epsilon||e^{-1})} \right]^N
= e^{-ND(e^{-1}+\epsilon||e^{-1})},
\end{IEEEeqnarray*}
which concludes the proof.
\end{IEEEproof}

\begin{IEEEproof}[Proof of Lemma~\ref{lemma:A_N}]
To prove the lemma, we will relate the sum $A_m=\sum_n n^{-\beta}$ with the corresponding integral, which can be evaluated as a closed-form expression.

Let $f$ be any \emph{decreasing} function defined on the interval $[k,l]$ for some integers $k$ and $l$.
Then, we can bound the integral of $f$ by
\[
\sum_{n=k+1}^l f(n) \le \int_k^l f(x)\mathrm{\,d}x \le \sum_{n=k}^{l-1} f(n).
\]
Rearranging the inequalities, we get the equivalent statement that
\begin{equation}
\label{eq:sum-bound}
f(l) + \int_k^l f(x)\mathrm{\,d}x
\le \sum_{n=k}^l f(n)
\le f(k) + \int_k^l f(x)\mathrm{\,d}x.
\end{equation}

Recall that $A_m=\sum_{n=1}^m n^{-\beta}$.
Thus we can apply \eqref{eq:sum-bound} with $f(x)=x^{-\beta}$, $k=1$, and $l=m$.
Since we know that
\[
\int_1^m x^{-\beta}\mathrm{\,d}x
= \frac{m^{1-\beta}-1}{1-\beta},
\]
this implies, using \eqref{eq:sum-bound},
\begin{IEEEeqnarray*}{rCcCl}
m^{-\beta} + \frac{m^{1-\beta}-1}{1-\beta}
&\le& A_m
&\le& 1 + \frac{m^{1-\beta}-1}{1-\beta}\\
\frac{m^{1-\beta}-1}{1-\beta}
&\le& A_m
&\le& \frac{m^{1-\beta}}{1-\beta},
\end{IEEEeqnarray*}
which concludes the proof.
\end{IEEEproof}

\begin{IEEEproof}[Proof of Lemma~\ref{lemma:pam-n-requests}]
Recall that $u_n(c)$ is a Poisson random variable with mean $\rho dp_n$.
For $n\le N_1$, we have $dp_n\ge p_1(\log d)^2$.
Therefore, using Lemma~\ref{lemma:chernoff}, we have for $n\le N_1$
\[
\Pr\left\{ u_n(c) > \left( 1+\frac{p_1}{4} \right)dp_n \right\}
\le e^{-\Omega(dp_n)}
\le e^{-\Omega( (\log d)^2 )}.
\]

For $N_1<n\le N_2$, we have $dp_n<p_1(\log d)^2$.
Define $\tilde u$ to be a Poisson variable with parameter $\rho p_1(\log d)^2 > \rho dp_n$.
By noticing that the function $\lambda\mapsto\Pr\{X(\lambda)>a\}$ defined on $\lambda\in[0,a)$, where $X(\lambda)$ is a Poisson variable with parameter $\lambda$, is an increasing function of $\lambda$, we have
\[
\Pr\left\{ u_n(c) > 2p_1(\log d)^2 \right\}
< \Pr\left\{ \tilde u > 2p_1(\log d)^2 \right\}.
\]
Lemma~\ref{lemma:chernoff} then upper-bounds this by $\exp\{-\Omega( (\log d)^2 )\}$.

The lemma then follows from a union bound over all the files.
\end{IEEEproof}

\begin{IEEEproof}[Proof of Lemma~\ref{lemma:mlp-prob}]
Since the MLP policy matches requests to caches starting from the least popular files, we first focus on requests for files less popular than file $W_{N_2}$.
Because the files follow a Zipf distribution, we can write
\(
p_{N_2} = p_1N_2^{-\beta} = p_1/{d^{(\beta+1)/2}}
\).

Every file less popular than $N_2$ is stored at most once across all the caches in the cluster.
Therefore, under MLP, a request for a file $n>N_2$ will remain unmatched only if the cache storing that file is matched to another request for another file $n'>N_2$.
Under the KS placement policy, the cumulative popularity of all files no more popular than file $N_2$ stored on a particular cache is less than
\[
p_{N_2} + p_{N_2+d} + p_{N_2+2d} + \cdots = O\left( p_1 d^{-\frac{\beta+1}{2}} \right).
\]
Each unmatched request for file $n>N_2$ corresponds to the event that there are at least two requests for the $M$ files less popular than $W_{N_2}$ stored on a cache.
Therefore, by the Chernoff bound (Lemma~\ref{lemma:chernoff}), the probability that a particular request for a file $n>N_2$ remains unmatched is at most $e^{-\Omega(d)}$.
By the union bound, the probability that at least one request for file $n\in\mathcal{R}$ such that $n>N_2$ is not matched by the MLP policy is at most $de^{-\Omega(d)}$.

Next, we focus on files $n\in\{2,\ldots,N_2\}$.
If the KS policy decides to store file $n$, it stores it on $w_n$ caches.
Therefore,
\ifdefined\isdraft
\begin{IEEEeqnarray*}{rCl}
\sum_{n=2}^{N_2} x_nw_n
&\le& \sum_{n=2}^{N_1} \ceil{ \left( 1+\frac{p_1}{2} \right)dp_n }
+ \sum_{n=N_1+1}^{N_2} \ceil{ 4p_1(\log d)^2 }\\
&\le& \sum_{n=2}^{N_1} \left[ \left( 1+\frac{p_1}{2} \right)dp_n + 1 \right]
+ \sum_{n=N_1+1}^{N_2} \left[ 4p_1(\log d)^2 + 1 \right]\\
&\le& \left( 1 + \frac{p_1}{2} \right)d(1-p_1)
+ 4p_1(\log d)^2 N_2
+ N_2\\
&\le& \left( 1 + \frac{p_1}{2} \right)d(1-p_1)
+ \Theta\left( d^{(1+1/\beta)/2} (\log d)^2 \right),
\end{IEEEeqnarray*}
\else
\begin{IEEEeqnarray*}{rCl}
\sum_{n=2}^{N_2} x_nw_n
&\le& \sum_{n=2}^{N_1} \ceil{ \left( 1+\frac{p_1}{2} \right)dp_n }
+ \sum_{n=N_1+1}^{N_2} \ceil{ 4p_1(\log d)^2 }\\
&\le& \sum_{n=2}^{N_1} \left( 1+\frac{p_1}{2} \right)dp_n
+ \sum_{n=N_1+1}^{N_2} 4p_1(\log d)^2 + N_2\\
&\le& \left( 1 + \frac{p_1}{2} \right)d(1-p_1)
+ 4p_1(\log d)^2 N_2
+ N_2\\
&\le& \left( 1 + \frac{p_1}{2} \right)d(1-p_1)
+ \Theta\left( d^{(1+1/\beta)/2} (\log d)^2 \right),
\end{IEEEeqnarray*}
\fi
which is less than or equal to $d$ for $d$ large enough.
Therefore, if files are stored according to KS Part 2, each cache stores at most one file from the set $\{2,\ldots,N_2\}$.

Let $\mathcal{K}_n$ be the set of caches storing file $n$ in a given cluster.
Let $E_{3,k}$ be the event that cache $k\in\mathcal{K}_n$ is matched to a user requesting a file $n>N_2$.
A cache will be matched to a user requesting a file less popular than $N_2$ only if at least one of the files that it stores, among those that are less popular than $N_2$, is requested at least once.
Since there are at most $M$ such files on each cache,
\[
\Pr\{E_{3,k}\} \le 1 - \left( 1 - O\left( d^{-(\beta+1)/2} \right) \right)^d.
\]
For a given constant $0<\epsilon<1$, there exists a $d(\epsilon)$ such that $\Pr\{E_{3,k}\}\le\epsilon$ for all $d\ge d(\epsilon)$.

For each file $n$, $N_1<n\le N_2$, let $E_{4,n}$ denote the event that more than $2p_1(\log d)^2$ of the $\ceil{4p_1(\log d)^2}$ caches in $\mathcal{K}_n$ are matched to users requesting some file $n'>N_2$.
By the Chernoff bound for negatively associated random variables~\cite{schrijver2003},
\(
\Pr\{E_{4,n}\} = \exp\{-\Omega( (\log d)^2 )\}
\).

From Lemma~\ref{lemma:pam-n-requests}, we know that with probability at least $1-Ne^{-\Omega( (\log d)^2 )}$, there are less than $(1+p_1/4)dp_n$ requests for each file $2\le n\le N_1$.
Therefore, with probability at least $1-Ne^{-\Omega( (\log d)^2 )}$, all requests for files in $\mathcal{R}$ such that $2\le n\le N_1$ are matched to caches by the MLP policy.

Finally, we now focus on the requests for the most popular file $W_1$.
Recall that if the KS policy decides to store this file, it will be stored on \emph{all} caches.
\addchange{If the total number of users is less than $d$ in every cluster,} then even if all the users requesting files other than $W_1$ are matched, the remaining caches can still be used to serve all requests for $W_1$.
\addchange{Thus the server only needs to broadcast $W_1$ if any cluster has more than $d$ users.
We can use Lemma~\ref{lemma:unmatched} to show that the expected total number of these excess users is $O(K^{-t_0})$, and so the expected rate required to serve file $W_1$ is $o(1)$.}
\end{IEEEproof}

\fi

\bibliographystyle{IEEEtran}
\bibliography{journal_abbr,caching}

\ifdefined\isextended
\else
\newpage

\begin{IEEEbiographynophoto}{Jad Hachem}
received the B.E. degree in Computer and Communications Engineering from the American University of Beirut in 2011, and the M.Sc. and Ph.D. degrees in Electrical Engineering from the University of California, Los Angeles, in 2013 and 2017 respectively.
Since 2017 he is with Google.
\end{IEEEbiographynophoto}

\begin{IEEEbiographynophoto}{Nikhil Karamchandani}
(S'05-M'12) received the M.S. degree from the Department of Electrical and Computer Engineering at the University of California at San Diego in 2007, and the Ph.D. degree in the Department of Electrical and Computer Engineering at the University of California at San Diego in 2011.
From 2011 to 2014, he was a postdoctoral scholar at the University of California at Los Angeles and at the Information Theory and Applications (ITA) Center at the University of California at San Diego.
He is currently an Assistant Professor in the Department of Electrical Engineering at the Indian Institute of Technology Bombay.
His research interests are in networks, communications, and information theory.
Dr. Karamchandani received the California Institute for Telecommunications and Information Technology (CalIT2) fellowship in 2005, the INSPIRE Faculty Fellowship in 2015, and the best paper award at COMSNETS 2017.
\end{IEEEbiographynophoto}

\begin{IEEEbiographynophoto}{Sharayu Moharir}
(S'10-M'17) received the B.Tech. degree in Electrical Engineering from IIT Bombay, Mumbai, India, in 2009, the M.Tech. degree in Communication and Signal Processing, and the Ph.D. degree in Electrical and Computer Engineering from The University of Texas at Austin.
She is an Associate Professor with the Department of Electrical Engineering, IIT Bombay.
Her research interests include algorithms and performance analysis for wireless networks and content delivery networks.
\end{IEEEbiographynophoto}

\begin{IEEEbiographynophoto}{Suhas N. Diggavi}
received the B. Tech. degree in electrical
engineering from the Indian Institute of Technology, Delhi, India, and
the Ph.D. degree in electrical engineering from Stanford University,
Stanford, CA, in 1998.  After completing his Ph.D., he was a Principal
Member Technical Staff in the Information Sciences Center, AT\&T
Shannon Laboratories, Florham Park, NJ. After that he was on the
faculty of the School of Computer and Communication Sciences, EPFL,
where he directed the Laboratory for Information and Communication
Systems (LICOS).  He is currently a Professor, in the Department of
Electrical Engineering, at the University of California, Los Angeles,
where he directs the Information Theory and Systems laboratory.

His research interests include wireless network information theory,
wireless networking systems, network data compression and network
algorithms; more information can be found at \textsf{http://licos.ee.ucla.edu}.
He has received several recognitions for his research including the
2013 IEEE Information Theory Society \& Communications Society Joint
Paper Award, the 2013 ACM International Symposium on Mobile Ad Hoc
Networking and Computing (MobiHoc) best paper award, the 2006 IEEE
Donald Fink prize paper award.  He is currently a Distinguished
Lecturer and also serves on board of governors for the IEEE
Information theory society. He is a Fellow of the IEEE.

He has been an associate editor for IEEE Transactions on Information
Theory, ACM/IEEE Transactions on Networking, IEEE Communication
Letters, a guest editor for IEEE Selected Topics in Signal Processing
and in the program committees of several IEEE conferences. He has also
helped organize IEEE conferences including serving as the Technical
Program Co-Chair for 2012 IEEE Information Theory Workshop (ITW) and
the Technical Program Co-Chair for the 2015 IEEE International
Symposium on Information Theory (ISIT). He has 8 issued patents.
\end{IEEEbiographynophoto}

\fi

\end{document}